# Development and Assessment of a Nearly Autonomous Management and Control System for Advanced Reactors

Linyu Lin, Paridhi Athe, Pascal Rouxelin, Maria Avramova, Nam Dinh
Department of Nuclear Engineering
North Carolina State University, Raleigh, NC
llin@ncsu.edu; pathe@ncsu.edu; pnrouxel@ncsu.edu; mnavramo@ncsu.edu; ntdinh@ncsu.edu

Abhinav Gupta
Department of Civil, Construction, Environmental Engineering
North Carolina State University, Raleigh, NC
agupta1@ncsu.edu

Robert Youngblood
Idaho National Laboratory, Idaho Falls, ID
robert.youngblood@inl.gov

Jeffrey Lane
Zachry Nuclear Engineering, Inc., Cary, NC
LaneJW@zachrygroup.com

**Abstract**

This paper develops a Nearly Autonomous Management and Control (NAMAC) system for advanced reactors. The development process of NAMAC is characterized by a three layer-layer architecture: knowledge base, the Digital Twin (DT) developmental layer, and the NAMAC operational layer. The DT is described as a knowledge acquisition system from the knowledge base for intended uses in the NAMAC system. A set of DTs with different functions is developed with acceptable performance and assembled according to the NAMAC operational workflow to furnish recommendations to operators. To demonstrate the capability of the NAMAC system, a case study is designed, where a baseline NAMAC is implemented for operating a simulator of the Experimental Breeder Reactor II during a single loss of flow accident. When NAMAC is operated in the training domain, it can provide reasonable recommendations that prevent the peak fuel centerline temperature from exceeding a safety criterion.

## 1. Introduction

With the advancement in computer performance, machine learning, and digital systems, interest in development of autonomous control systems has increased in a variety of fields from industrial manufacturing to unmanned space, ground vehicles, and nuclear reactors. Autonomous control systems are intelligent systems with self-governance ability to perform and execute control functions in the presence of uncertainty for an extended time [1]. The degree of autonomy of an autonomous control system depends upon the extent to which it can perform fault diagnosis, planning, forecasting, and decision-making under uncertainty, without human intervention [2].

Owing to the inherent risk and uncertainty associated with the operation of nuclear reactor systems, the design of autonomous control systems is a challenging task. Over the past several years, different techniques have been adopted to develop functions related to autonomous control and operation of nuclear reactor systems. Upadhyaya *et al.* [3] [4] developed an autonomous control system for a space reactor system (Fast spectrum Lithium cooled reactor) with Model Predictive Control (MPC) using a Genetic Algorithm for optimization. Fault detection in this system is performed using Principal Component analysis. Cetiner *et al.* [5] developed a Supervisory Control System (SCS) that uses a probabilistic decision-making approach using fault tree and event tree in conjunction with deterministic assessment of plant state variables for autonomous control and maintenance of advanced small modular reactors. Groth *et al.* [6] [7] use dynamic Probabilistic Risk Analysis (PRA) for fault detection and management, and counterfactual reasoning for decision analysis in a Sodium fast reactor during earthquake-induced transients. This system is called the Safely Managing Accidental Reactor Transients (SMART) system. Lee *et al.* [8] developed an autonomous operation algorithm for core damage prevention (loss of coolant accident, Steam generator tube rupture) in a Pressurized Water Reactor. This work uses a Function-based Hierarchical Framework (FHF) and an advanced Artificial Intelligence (AI) algorithm like Long Short-Term Memory (LSTM) for plant state diagnosis and control. All the autonomous control systems discussed here perform diagnosis and decision-making for fault management and control. In the case of the SCS and SMART systems, a strategic decision analysis is performed based on the consequence of decision choices while in the case of the space reactor system and FHF, decision making is implicitly performed based on some preset decision preferences.

In this study, a Nearly Autonomous Management and Control (NAMAC) system is designed to provide recommendations to the operator for maintaining the safety and performance of the reactor. The development of the NAMAC system is based on three elements:

- Knowledge base – a class of databases, scenarios, and models to support the control and risk management of the reactor;
- Digital Twin (DT) – a knowledge acquisition system to support different NAMAC functions (i.e., diagnosis, strategy planning, prognosis, strategy assessment, etc.);
- Operational workflow – an assembly of DTs to support operator's decision-making or to make direct operational recommendations.

Comparing to the reviewed intelligent systems, NAMAC is a computerized safety case that aims to achieve an alignment of NPP safety design, analysis, operator training, and emergency management by furnishing recommendations to operators for effective actions that will achieve particular goals, based on the NAMAC's knowledge of the current plant state, prediction of the future state transients, and reflecting the uncertainties that complicate the determination of mitigating strategies. Such knowledge is extracted from the knowledge base by machine-learning algorithms and stored in DTs of various functions. Although this is not the first time machine learning algorithms or DTs are used in the autonomous control system, this is the first time that these tools are implemented and combined in a system with a structured workflow in order to promote rigor, comprehensiveness, and realism in safety cases. Moreover, NAMAC system recognizes the importance of explainability by deriving recommendation from an intelligible technical basis. This includes an argument-based operational workflow, operating procedures and reactor technical specifications in the knowledge base, and a modular NAMAC architecture with DTs of different autonomous functions. Key DTs in NAMAC include:

- Diagnosis – Monitors safety significant factor(s) based on observed sensor data;
- Strategy inventory – Identifies feasible control options based on plant state diagnosis, safety and control limits;
- Prognosis – Forecasts plant state for each control option;
- Strategy assessment – Ranks the control options based on the consequence and user defined preference structure related to safety, operability and/or performance of the reactor.

To evaluate and demonstrate the capability of DTs and the NAMAC system, a case study is designed for the control of Experimental Breeder Reactor – II (EBR-II) during a single Loss Of Flow Accident (LOFA). To further evaluate the scalability and uncertainty of DTs and NAMAC, test cases are designed with different sources of uncertainty. For DTs tests, sources include input data, model fits, scope compliance, etc., while for NAMAC tests, accident scenarios lying outside the training domain are used. To avoid severe consequences due to NAMAC uncertainty, a global discrepancy checker is implemented to determine if the plant is moving towards the expected system state after the control actions are injected. If the discrepancy between expected and observed states exceed a limit, an anomaly is claimed, and the operator is alerted. Meanwhile, a safety-oriented control action, i.e. SCRAM, is recommended.

The organization of the paper is as follows. Section 2 describes the concepts of three-layer hierarchical NAMAC development process, DT technology and its implementation in NAMAC, plant simulator model, data generation engine, and NAMAC operational workflow. Section 3 presents a case study where NAMAC is controlling the simulator of EBR-II during a single LOFA based on DT implementations and operational workflows. Moreover, tests are performed to evaluate the uncertainty of DTs and NAMAC systems with different sources of uncertainty. Section 4 presents conclusions of this study.

## 2. Nearly Autonomous Management and Control (NAMAC) System and Design

This section describes the NAMAC architecture, the development of the DTs, and the construction of the knowledge base. The first part of this section illustrates the three-layer NAMAC architecture, including knowledge base, DT development process (for different NAMAC functions), and NAMAC operational workflow. The second part of this section illustrates the concept of DTs and how they are implemented in the NAMAC system. The third part of this section describes the GOTHIC plant simulator and data generation engine. The fourth part of this section describes the NAMAC operational workflow based on the assembly of different DTs.

## 2.1. NAMAC Architecture

The development process of NAMAC can be demonstrated by the three-layer architecture (see Figure 1). The three layers are: (i) Knowledge base, (ii) NAMAC developmental layer, and (iii) NAMAC operational layer.

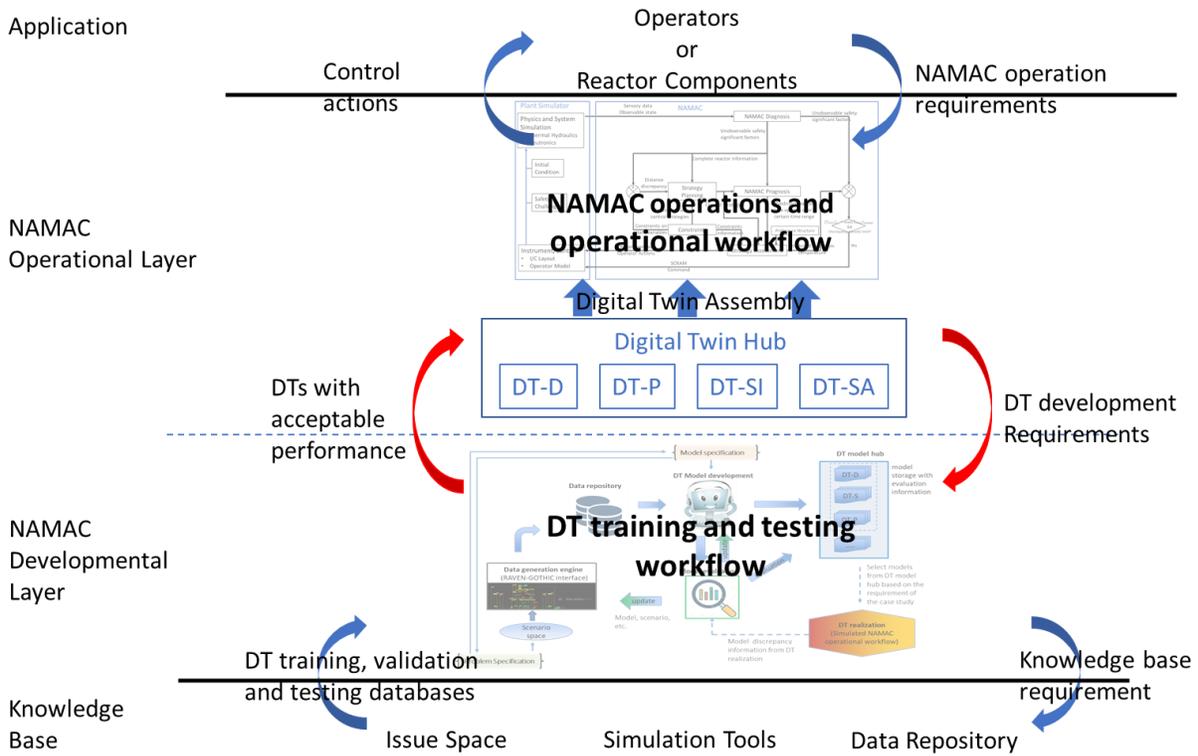

Figure 1: NAMAC integral workflow

The first layer focuses on development of the knowledge base. The knowledge base serves as the foundations to the whole scheme, and the upper layers store or make use of partial information from the knowledge base. The knowledge base for NAMAC development can be classified into three components: (i) issue space, (ii) simulation tool, and (iii) data repository. Issue space defines the scenario in mathematical formulations. The simulation tool (system code with adequate fidelity) is required for generating training/testing data set for the development of different DTs. In this work, we employ GOTHIC [9] for system level simulations and RAVEN [10] as the sampling tool. Data repository has two elements: knowledge element and data element. Knowledge element consists of literature or information related to operating procedures and training materials, system configuration, initial conditions, reactor failure modes, experimental data, benchmarking results, etc. Data element consists of data generated by the simulation tool for development of NAMAC DTs and plant data collected from operational histories, transients, and events.

The second layer focuses on the development and implementation (training and testing) of NAMAC DTs. These DTs can be treated as surrogate models for specific sets of information and knowledge from the knowledge base. For example, Digital Twin for Diagnosis (DT-D) aims to monitor the unmeasurable and unobservable state variables by storing correlation and dependencies among different state variables; Digital Twin for Prognosis (DT-P) is used to predict the short-term transient and the consequences of control actions by storing the time-propagation of state variables with respect to reference information, including initial conditions, control actions, histories, etc. More details about DTs training and testing are discussed in section 2.2.

Once DTs are implemented, they are fed to the third layer and assembled based on the operational workflow. Since NAMAC aims to support operators' decision-making by making recommendations, a logic scheme is needed to support the final recommendation based on real-time observations and records by DTs. For example, NAMAC needs to figure out the complete states of the reactor by monitoring the unobservable state variables with DT-D; NAMAC also needs to understand the consequences of control actions by predicting the short-term transient of state variables with DT-P. More details about the design and implementation of NAMAC operational workflow is discussed in section 2.5. The operational layer also aims to test the performance of NAMAC by coupling the system with a nuclear reactor or a plant simulator.

The NAMAC structure discussed here is highly modular. We believe that this has important advantages in scalability and interpretability, which will become more important when a much broader issue space is considered. Moreover, the modular architecture allows for a plug-and-play character such that NAMAC can be more adaptive to different reactor designs, instrument and control system, hardware platform, etc.

To avoid major failures during DT training and NAMAC operations, discrepancy checkers are incorporated between layers. The discrepancy checker between development and operational layers aims to determine if the DT has satisfied the acceptance criteria with respect to accuracy and coverage, where coverage is defined based on the comparison between training and application scenarios. The discrepancy checker in operational layer aims to continuously monitor the reactor states and to compare them against those from DTs and NAMAC. If the observed states largely deviate from DTs' output, it is suggested that the DT outputs and NAMAC recommendations are not trustworthy either because they are operated outside the training domain or their accuracy are not high enough in the developmental stage. To avoid the situation where NAMAC overlooks the failure conditions due to inherent errors, the operator will be alerted and suggested to take safety-minded control actions, i.e. SCRAM, when the discrepancy is larger than a criterion. It is stressed that NAMAC does not necessarily intend to replace traditional control and management guidelines, although this might be an option for microreactors operating remotely. NAMAC talks to the operators about what is likely to work best; it is intended to be useful whenever operators are required to monitor numerous and diverse systems in order to take the proper action.

Other architectures, including FHF and supervisory control architecture, can be found in works by Lee *et al.* [8] and Cetiner *et al.* [11]. FHF decomposes the control of complex reactor system into three levels: the goal, function, and system levels. The FHF starts at the conceptual level with Nuclear Power Plant (NPP) safety goals that must be achieved by the autonomous system to ensure the reactor and public safety, for example, to prevent core damage and the release of radiation, to mitigate the consequences of accidents, etc. The function level includes functions that are designed to achieve the goal, including reactivity control, containment integrity, reactor coolant system, etc. The system level identifies the system, components, and input/output parameters of control systems that are designed to satisfy the safety functions. The supervisory control architecture also has goal-function-system structure, but it further decomposes the function level, into decision, prognostic, and diagnostic sub-levels. The objective is to achieve complete and accurate understanding in the reactor interval states with a successive delegation of duties. For example, the decision-makings depends on prognostic estimation of remaining service lift of key components, which further depends on the diagnostics of components' conditions. In NAMAC, the three-layer structure is a natural extension to the Data-Information-Knowledge-Wisdom (DIKW) pyramid, which is usually treated as a computational representation of intelligence system [12]. Such an architecture is claimed to be able to maximally ensure system adaptability when complexity grows [13].

## 2.2. Digital Twins

This section is divided into five parts. The first part defines DT technology based on the available materials. The second and third parts illustrate the requirement and design of DTs with data-driven methods. The fourth part introduces one of the data-driven modeling techniques, feedforward neural network, for developing different DTs. The fifth part illustrates the phases of DT development based on their maturity with respect to reactor application. It is stressed that the current development is at a scoping stage, where the major objectives are to learn from experience and to mature the framework. It is argued that the general applicability of DT definition and the supreme expressive power of neural networks make DT technology by ML algorithms largely applicable to different system designs.

### 2.2.1. Digital Twin Technology

A DT is a digital representation of a physical object or system that relies on real-time and past-history data to evaluate its complete states, to predict future transients, and to recommend appropriate control actions. During the operation of a NPP, DTs can help the operator to sustain an accurate understanding of the physical system, improve the effectiveness of control, and avoid human errors. Considering the complexity of physical systems like nuclear reactors, the implementation of DTs requires diverse technologies, including modeling and simulation, data analytics, AI, etc. To guide the development of DTs, Kahlen *et al.* [14] define DT based on three attributes: digital twin prototype (DTP), digital twin instance (DTI), and digital twin environment (DTE).

The DTE represents the application space for operating the DTs, and it can be further classified into predictive and interrogative DTEs. The predictive DT, sometimes known as prognosis DT, is used for predicting future behavior, and the predictive performance starts from the current point in the product's lifecycle at its current state, and moves forward in time. The interrogative DT, also known as diagnosis DT, is used to interrogate the current and past histories of certain objects or systems.

The DTP contains models and tools necessary to describe and produce a virtual version that duplicates or twins the physical version. Generally, there are three classes of options: model-free methods, model-based methods, and reasoning-based methods. The parameters of a model-based method have a physical or qualitative meaning, and the number of parameters is usually fixed. Classical model-based methods for diagnosis include parity equations, diagnostic observer, Kalman filters, and parameter estimation. For model-free methods, the parameters represent the joint distribution or statistical hypotheses concerning the behavior of observed random variables [15]. Calibrations are needed to estimate these parameters based on experimental measurements or databases, and they can be sampled from probability distributions. Reasoning-based models are usually implemented by if-then-rules, and tools include Answer Set Programming, Fuzzy Logics, Bayesian networks, Prolog, etc. They have been successfully used mainly in interrogative DTEs.

DTI refers to a specific instance of an individual DT that is directly linked to a physical object or system. In applications, DTI usually refers to the scenarios, I/O, and user interfaces of DTs. It is stressed that compared to DTP, the DTI contains more details since the DTP is designed to reflect multiple or a sequence of instances. Meanwhile, the DTI needs to be more specific such that it can be clearly defined for a specific scenario, initial/boundary condition, time point, user interface, etc.

In this study, the DTE is limited to applications of system controls for advanced NPPs and referred to as DT functions. To be specific, the interrogative (diagnosis) DTE is used to monitor state variables that are safety significant to the System Thermal-Hydraulics (STH) in the primary loop of an advanced reactor, while the predictive (prognosis) DTE is used to predict the transients of selected state variables for a designated time period. Additionally, DTs for strategy inventory and for strategy assessment are introduced to support the decision-making process in NAMAC. Moreover, this study limits the scope of DTP to model-free for modeling the DTs.

2.2.2. *Digital Twins in NAMAC*

DTs establish the basis for NAMAC operation by learning from the knowledge base. DTs in NAMAC are developed as knowledge acquisition system that consist of various surrogate models to support different NAMAC functions (e.g., diagnosis, strategy inventory formulation, prognosis, strategy assessment). As NAMAC adopts a modular architecture, DTs are designed and developed as individual modules based on their functions in NAMAC system. Depending on these functions, the DT in NAMAC consists of four categories:

- Interrogative DT or DT for Diagnosis (DT-D),
- DT for Strategy Inventory (DT-SI),
- Predictive DT or DT for Prognosis (DT-P),
- DT for Strategy Assessment (DT-SA).

The workflow for DT development is shown in Figure 2. The problem specification defines the analysis purpose, quantities of interest, issue space (or scenario space), system configurations, failure modes, etc. This information is gathered from the knowledge base (layer 1 in NAMAC integral workflow). Based on the issue space, model specifications for each DT are determined. The database for training different DT models is generated by sampling the scenario space using the data generation engine (RAVEN-GOTHIC interface). Since the type and class of DTs vary, model specifications are defined based on functional, modeling, and interface requirements for each class of models. The functional requirement states the function of each DT model (constituting DTE). The interfacing requirement describes the inputs and outputs of the model (constituting DTI). The modeling requirement determines the implementation methods of the model (constituting DTP).

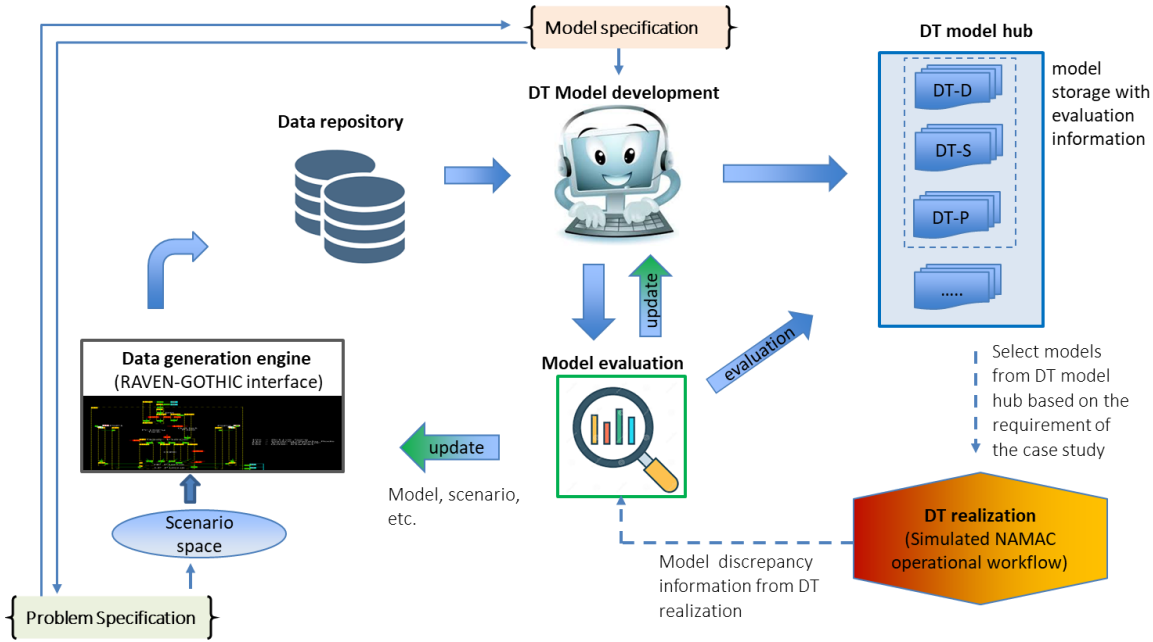

Figure 2: DT developmental workflow

### 2.2.3. Implementation of DTs in NAMAC

The DT-D aims to assimilate data from the operating plant to evaluate the complete states of a physical systems, including unobservable state variables, operating and fault conditions. In nuclear applications, the interrogative DTs are also known as fault detection and diagnosis (FDD) models. Since the model-based FDD methods are limited due to the requirement for an accurate model, model-free FDD (especially the data-driven FDD methods) have become popular as it does not explicitly rely on the mathematical models. By constructing DT-D with machine learning and AI approaches, significant progress has been made with model-free methods for FDD applications in reactors. As one of the model-free methods, data-driven methods have been applied with various tools, including artificial neural networks, support vector machine, auto-associate kernel regression, Principal Component Analysis, etc. The objective of these approaches is to construct a relationship between correlated state variables. If all variables are observable, by monitoring the changes in residuals between the measurements and output variables from DT-D, system faults can be detected when there are statistically abnormal changes. If outputs are unobservable safety-significant state variables, the DT-D can be used to support predictions by DT-P, to find appropriate control actions from the operational manuals and procedures, or to directly inform operator. In this study, DT-D is used to determine the unobservable state variables that are potentially safety-significant, defined as Safety-Significant Factors (SSFs), including peak fuel centerline temperature and peak cladding temperature. Meanwhile, NAMAC is making decisions and recommendations based on these parameters, which are related to both the performance and safety conditions of the reactors. They are unobservable as this study assumes that the SSFs cannot be directly measured by sensors. Eq. 1 shows a general form of DT-D; $f_{DT-D}$ is the data-driven model of DT-D; $X_D = [x_1, x_2, \ldots, x_i, \ldots]$ represents input variables of the DT-D model; $P_D = [p_1, p_2, \ldots, p_j, \ldots]$ represent training and design hyper-parameters of DT, including network structures, error requirements, etc.; $KB_D$ represents the knowledge base used to train the data-driven models, including simulation databases, user knowledge in feature selections and preprocessing; $\varepsilon_D$ represents the error of DT-D model. Although parameters $P_D$ and $KB_D$ are implicitly involved, they are known to have significant impacts on the data-driven models and machine learning algorithms [16].

$$SSF(t) = f_{DT-D}(X_D, P_D, KB_D) + \varepsilon_D \qquad \text{Eq. 1}$$

Figure 3 shows the principal scheme of the data-driven DT-D. The dashed line represents an indirect or implicit involvement of the DT-D model.

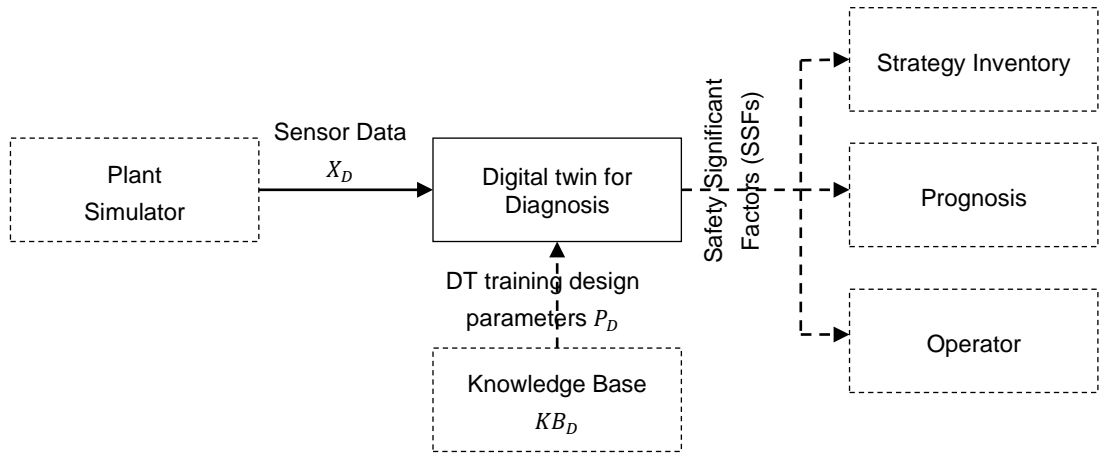

Figure 3: Scheme of data-driven DT-D model for evaluating SSFs.

Besides data-driven methods, signal-based FDD methods are also classified as another model-free method, where decisions are made by comparing features from signal against "baseline" characteristics. Spectrum analysis, and its extensions, time-frequency analysis and wavelet transform, are typical signal-based FDD methods. Signal-based methods have been used in NPP applications for instrument monitoring, equipment vibration monitoring, acoustic emission monitoring, etc. Compared to data-driven methods, signal-based FDD methods are mostly limited to component and sub-system levels. To recover the complete conditions of reactor or suggest control actions to operators, data-driven methods tend to be more attractive for practical applications since they are more feasible to implement than mechanistic models. Moreover, data-driven methods are preferred since they can learn from a large amount of operational and maintenance data, and they are able to capture complex correlations among state variables. However, the applicability of data-driven methods is mostly limited by the training and design hyper-parameters ($P_D$) and the training knowledge base ($KB_D$) of DTs. Therefore, in addition to the regular testing procedures during the training of data-driven models, assessments are needed in the context of application scope.

DT for strategy inventory (DT-SI) aims to identify available actions based on the current state of the reactor (i.e., diagnosed plant state at the current time step), reactor's safety limits, and component's control limits. Implementation of strategy inventory requires knowledge of reactor system dynamics, technical specification, operating procedure, etc. Different approaches can be adopted for implementation of DT-SI. The basic option is to use a grid-based algorithm that search for available actions across the entire action space. Another option is to use AI declarative approaches like Answer Set Programming (ASP). Hanna *et al.* [17] have used ASP to build an operator support system that can perform fault diagnosis, informs operator of different scenarios and consequence, and generates control options. Reinforcement Learning (RL) can also be used to generate control options. RL is an intrusive approach that requires a model of the reactor ("environment" in RL terminology) to explore control actions based on the reward function. In the supervisory control system by Cetiner *et al.* [5], control options are determined based on a probabilistic decision-making module and fault tree and event tree (FT/ET) models. However, the amount of data points in the current knowledge base is too limited to ensure the trustworthiness of strategies made by RL. In addition, since the present case study aims to demonstrate NAMAC with a simplified issue space, FT/ET is not available for strategy inventory. At the same time, ASP is suitable for problems with complex logics and reasoning. Since the current strategy inventory adopts a simple reasoning process, it greatly decreases the necessity of applying ASP. As a result, DT-SI is implemented by grid-based sampling methods. Although the searching process can be computationally expensive than other sophisticated techniques, it is more preferred to identify all possible actions with simple techniques.

Figure 4 shows the scheme of DT-SI model to determine a feasible set of control actions according to the diagnosed state of the reactor (complete reactor information) at the current time. The knowledge base for DT-SI provides the success path for mitigative actions and constraints due to control and safety limits of the reactor components. The dashed line represents an indirect or implicit involvement of the DT-SI model.

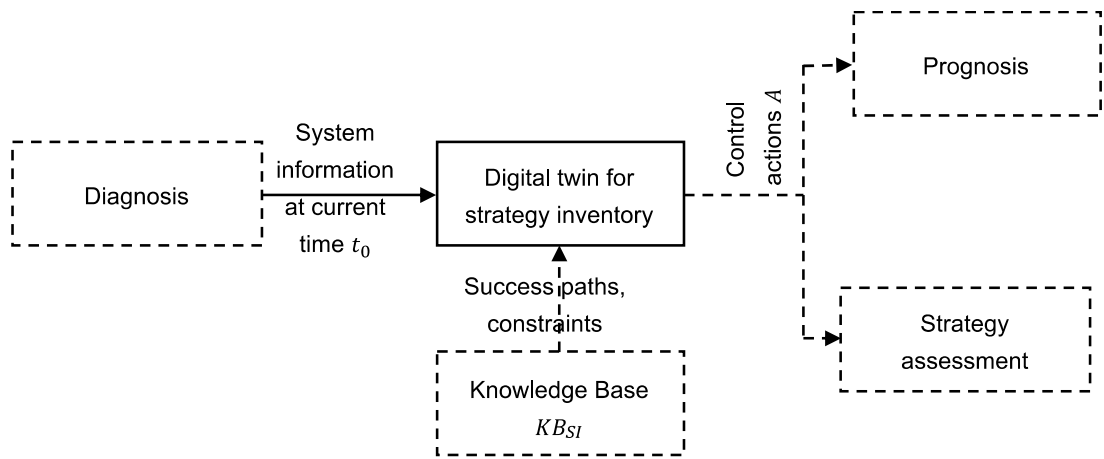

Figure 4: Scheme of DT-SI model to determine a feasible set of control action based on the plant diagnosis.

The DT-P aims to predict the future transients of state variables or lifecycles of certain components based on the past histories and current information. In manufacturing and medical applications, they are usually implemented through data analytical models [18]. In nuclear applications, most predictive DTs are implemented by combining model-based methods with a Probabilistic Risk Analysis (PRA) framework: the PRA framework accounts for the component failures and control actions, while the system code predicts consequences along all paths. In the supervisory control system by Cetiner *et al.* [5], event/fault trees are combined with the MODELICA plant model to predict the future transients with faults reflected by PRA. In "SMART" procedures, a dynamic Bayesian network is constructed to represent the causal relationships between previous system states and likely states of reactor parameters in the event of instrument or control-room failures. There are a few applications under development for constructing DT-P with data-driven methods. In the operator support tool by J.H. Lee et al. [7], deep learning techniques are trained as a surrogate of analysis toolkit for dynamic accident progression. The analysis toolkit is built by combining dynamic event tree with a set of simulation codes. In the SP-100 system [4], a Controlled Auto-Regressive and Integrated Moving Average model is constructed, and its parameters are calibrated continuously based on measurement values. Compared to prognosis by direct plant models, data-driven models can make predictions with acceptable accuracy starting from an arbitrary point during a transient. Moreover, the running speed of DT-P by data-driven methods is fast enough to support real-time operations. The performance of dynamic Bayesian network usually relies on expert inputs in assigning the conditional probabilities and connections between different nodes, while the data-driven methods can capture time sequences and nonlinear correlations among variables by directly learning from the database. As a result, this study trains a synthetic DT-P model with data-driven methods to predict the consequence factor $C$ of control actions $A = \{a_1, a_2, \ldots, a_i, \ldots\}$ and information $X$ at the current time point $t_0$. Eq. 2 shows the general form of DT-P, and $f_{DT-P}$ represents the DT-P model implemented by the data-driven methods; $\varepsilon_P$ represents the error of DT-P model. Similar to the DT-D, model parameters $P_P$ and knowledge base $KB_P$ are also included.

$$C_{A,X} = f_{DT-P}(A, X_{t_0}, P_P, KB_P) + \varepsilon_P \qquad \text{Eq. 2}$$

Figure 5 shows the principal scheme of the data-driven DT-P, and the dashed line represents an indirect or implicit involvement to the DT-P model. By including a decision-analysis model (DT-SA), the consequence factor $C$ predicted by DT-P can be used to assess strategies from the strategy inventory. Meanwhile, the factor $C_{A,X}$ can also be used directly by operators for deciding control actions.

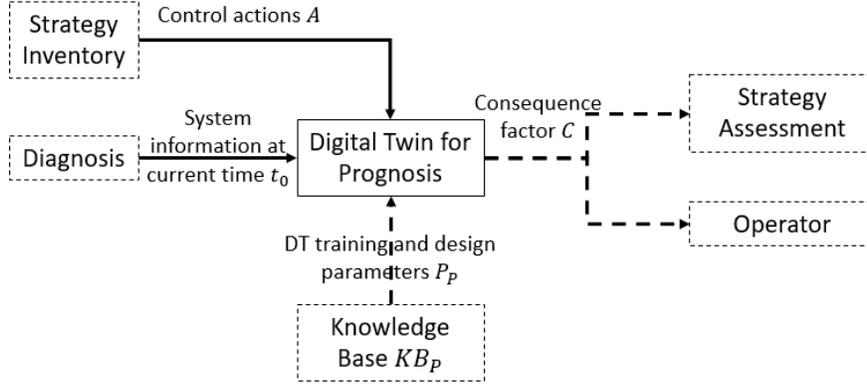

Figure 5: Scheme of data-driven DT-P model for predicting the consequences of control actions based on the current system information.

The DT for Strategy Assessment (DT-SA) can be described as a decision-making module that aims to rank the control actions based on a consequence factor (obtained from prognosis) and a user-defined preference structure. The preference structure is based on decision-makers' preference for safety, operability and/or performance of the reactor. Different approaches can be adopted for implementation of DT-SA. Apostolakis *et al.* [19] use multi-attribute utility theory to evaluate decision choices for reduction in the containment integrated leakage rate tests. The decision choices are evaluated based on relative preference for safety, economics, and stakeholder relation. In the supervisory control system by Cetiner *et al.* [5], a utility theory algorithm is employed to perform strategy assessment. Utility functions are defined for key reactor parameters based on the safety and operation limits. In this study, strategy assessment is performed by finding the limit surface [20] where the control inputs cause transition from safe to unsafe consequences. Control options are ranked based on the safety margin, while impact of component reliability and operability (economic aspect) is not included in the initial demonstration documented in this paper but will be considered in the future work. Hence, preference structure is governed by the safety criteria alone. Figure 6 shows the scheme of DT-SA for strategy assessment based on the preference structure. The knowledge base for DT-SA provides necessary information and/or guidance to construct the preference structure for decision making. The dashed line represents an indirect or implicit involvement of the DT-SA model.

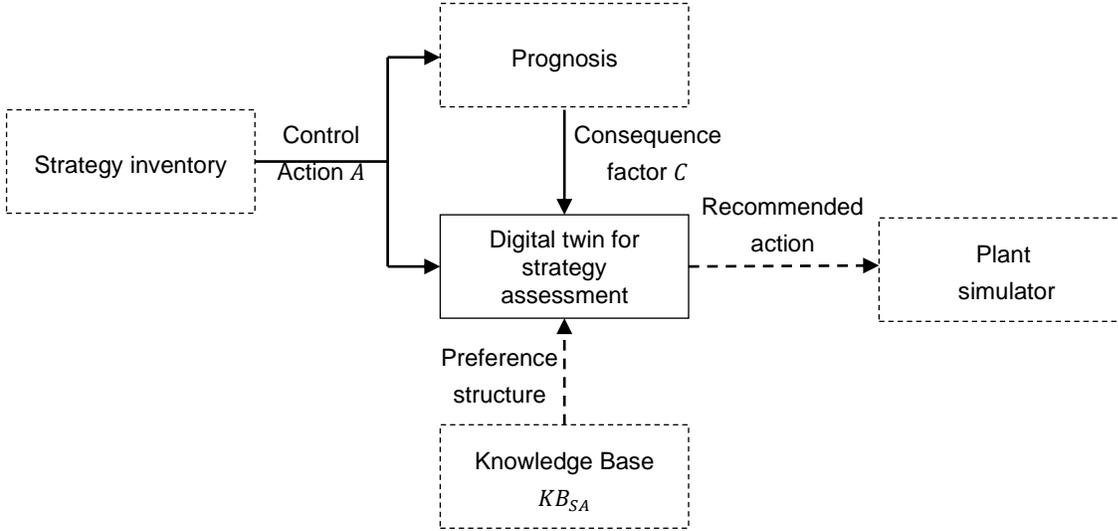

Figure 6: Scheme of DT-SA for strategy assessment according to the preference structure

*2.2.4. Data-driven modeling for digital twins in NAMAC*

The mathematical expressions of data-driven models typically have implicit model forms and model parameters with no engineering or physical meanings. When constructing DTs for diagnosis and prognosis, the data-driven models use processed data from simulations and/or operations to extract information out of big data sets. Meanwhile, preprocessing techniques filter the raw data to improve the training efficiency and predictive capability. Popular data-driven techniques include polynomials, artificial neural networks, principal component regression, and support vector regression, etc. to capture the relevant process behavior from data. According to the difference in calibration or learning approaches, there are supervised and unsupervised learning algorithms. For supervised learning, inputs and outputs are clearly labelled, and the goal is to calibrate the model parameters such that the errors between predicted and measured outputs are minimized. An unsupervised algorithm does not use output data and no reference values are needed for model calibration. An example of unsupervised learning is clustering or principal component analysis.

In this study, one of the supervised algorithms, feedforward neural network, is trained for the implementation of DT-D and DT-P. The topology of feedforward neural networks consists of a set of neurons connected by links in a number of layers. A multilayer feedforward network is shown in Figure 7. The basic configuration of feedforward networks usually includes an input layer, hidden layers, and an output layer. After the training, the same mapping with fixed weights are applied from the input to the output space. In other words, the state of any neuron only depends on the input-output pattern, but not the initial and past states of the neuron. Therefore, no dynamic representation is involved, and such network topology is classified as static neural networks. A static feedforward network can be built with simple optimizing algorithms, and it has become one of the most popular network architectures in use today.

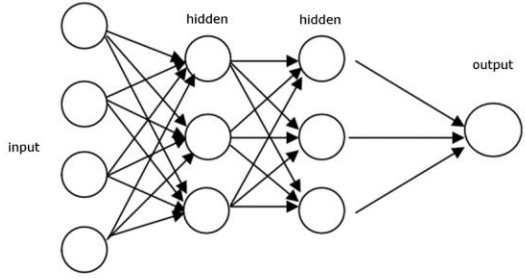

Figure 7: The basic configuration of a multi-layer feedforward neural networks

Eq. 3 to Eq. 5 show the mathematical formulation of feedforward neural networks. Eq. 3 gives the input vector $\vec{x}$ with $n$ elements. Eq. 4 shows the model inside each hidden unit ($HU$) where $i$ and $j$ are the $i^{th}$ inputs, and $j$ indexes hidden units. $w_{ji}$ and $b_j$ are weights and biases for each hidden unit. $\sigma$ is the nonlinear activation function. $\hat{y}$ is the output of neural networks, where $m$ represents the $m^{th}$ output layer ($o$).

$$\vec{x} = [x_1, x_2, \ldots, x_n]$$
Eq. 3

$$HU_j(\vec{x}) = \sigma\left(\sum_{i=1}^{n} w_{ji} x_i + b_j\right)$$
Eq. 4

$$\hat{y}(\overrightarrow{HU}) = \sum_{i=1}^{m} w_{oji} HU_i + b_o$$
Eq. 5

Eq. 6 is a loss function that optimizes parameters of neural networks, where $N$ is the total number of training targets $y$. For regression problems, the $L_2$ squared norm is usually used.

$$L_D = \frac{1}{2N} \sum_{i=1}^{N} (y_i - \hat{y}_i)^2$$
Eq. 6

To avoid overfitting, a regularization term is added to the loss function (Eq. 7), where $\alpha$ and $\beta$ are loss function parameters. $L_W$ is the sum of squares of the network weights. In Bayesian regularization, neural network parameters $\alpha$, $\beta$, and $w_{ji}$ are treated as random variables and updated according to the Bayes' rules [21].

$$L_F = \beta L_D + \alpha L_W$$
Eq. 7

The objective of a neural-network training is not only to have a model that could best represent the training data, but also new data on which the model will be used to make predictions. However, it is a challenge to generalize the model from known data points to new data points in unknown domains, and this challenge is a generalization problem [22]. In addition to a good-fit model that both learns and generalizes well, a model can be either under-fit or over-fit. Underfitting is easier to detect and mainly controlled by the final training loss and maximum epoch number in the training process. Since the stochastic gradient descent algorithm is an iterative learning algorithm that uses the training dataset to update the neural network model, the epoch number is a hyperparameter of gradient descent that controls the number of complete passes through the training dataset. The model will stop updating if the loss function $L_D$ reaches a final training loss or epoch number reaches the maximum epoch number. However, an over-fit model can hardly be diagnosed when only the training data is available since the model output usually has very small biases and variance comparing to the real data. Common techniques for reducing overfitting include setting early stop criteria, separating a validation set from the training data, testing the model with more data. Meanwhile, it is also suggested that the complexity of the network should be adaptive and comparable to the idiosyncrasies of training databases [23]. Therefore, considering the severe consequences of DT accuracy in engineering applications, it is stressed that both errors and generalization capability of any DT implementations should be assessed after the training. Meanwhile, neural-network-based DTs should evolve from simple structures (i.e. topologies and hyperparameters) to more complex patterns.

### 2.2.5. *Phases of data-driven digital twin development*

Considering the complexity of parameter selections and knowledge base construction, the DT development process is classified into three phases: scoping, refinement, and maturation. Table 1 defines the phases of DT training and developments for autonomous control systems. In all phases, it is required that DT prototypes and instances are implemented in the designed environments for the target applications. Meanwhile, it is also required that DT errors be analyzed, and that they satisfy the acceptance criteria. At the scoping stage, users design the acceptance criteria, DT training, and design hyper-parameters. However, there are potentially large uncertainties and biases due to the limited knowledge of user groups. Although the DT uncertainties satisfy the acceptance criteria, they are limited by users' knowledge in designing the evaluation metrics and acceptance criteria. As a result, at the scoping stage, it is likely that the DT uncertainty in the target applications is so large that it could alter outcomes of directly related blocks. For instance, the large uncertainty of DT-D model could lead the operator to misunderstand the states or faults in the reactor system; it may also change the prediction of consequence factors by the DT-P model. Many lessons will be learned at this stage, especially in evaluating the effects of hyper-parameters, designing assessment metrics, and determining potential technology impacts.

At the refinement stage, the major sources of uncertainty and bias are refined by improving users' knowledge or formalizing the development and assessment process. The DT uncertainty can still be large, however, they should be better characterized and continuously decreased in a consistent and transparent manner. Uncertainty quantification is one of the major techniques in the refinement stage that refers to the activity of identifying, understanding, and quantifying all possible uncertainties within the system of interest [24]. For ML algorithms, Bayesian inference can be used to optimize hyperparameters based on a prior distribution and additional observations [25]. Interval analysis is used to accelerate the convergence of ML to the global minimal point or to guarantee that all solutions are found to any degree of accuracy with guaranteed bounds [26]. Meta-learning, i.e. learning to learn, divides a large learning task into two hierarchies of learning: base learner and meta learner. Compared to the classical training for ML-based DTs, denoted as the base learner, there is also a meta model that optimizes the base learner by updating its parameters via a meta-knowledge base [27]. In addition to the separate DTs development and assessment, the quantitative software reliability method [28] and digital instrumentation & control (I&C) assessment [29] evaluate the reliability and risk of the autonomous control system and digital I&C by quantifying the software failure rates and the risks to reactor component and system. As a result, at the refinement stage, the impacts of uncertainty and bias on operators' decisions and directly-connected DTs are known, and resources can be directed to areas with insufficient knowledge. The largest amount of resources is expected in the refinement stage. A formalized framework for knowledge elicitation or parameter optimizations is required to improve the efficiency and effectiveness in resource allocations.

At maturation stage, major sources of uncertainty and bias are verified and optimized based on the best available methods and techniques. There are still uncertainties, but they are small compared to the licensing and regulatory requirements. Efforts at maturation stage focus on the license, certification, approval, and applications of DTs and NAMAC, including reactor controls, reactor system/sub-system/component designs, etc.

Table 1: Definition of phases in DT development

| Phase | Major Sources of Uncertainty and Bias | Consequence |
|---|---|---|
| 1. Scoping | The acceptance criteria are assigned by user(s) with limited knowledge. The DT training and design hyper-parameters are selected or assigned by user(s) with limited knowledge | DT uncertainty in the final code adequacy is large and their impacts on the operator decision and outcomes of directly-connected DTs are uncertain |
| 2. Refinement | The acceptance criteria are refined by user(s) with improved knowledge or formalized frameworks. The DT training and design hyper-parameters are refined by user(s) with improved knowledge or formalized optimization scheme | DT uncertainty is large, but its impacts on operator or directly-connected DTs are sufficiently characterized |
| 3. Maturation | The acceptance criteria are verified. The DT training and design hyper-parameters are optimized | DT Uncertainty is negligible compared to the defense-in-depth |

### 2.3. Plant simulator and data generation engine

For NAMAC testing and "proof of concept" demonstration, a plant simulator is required. A system code with embedded virtual instrumentation and control knobs is employed as a plant simulator. A simplified five-channel model for EBR-II has been developed in GOTHIC and benchmarked against experimental data to serve as a plant simulator [30]. This simplified EBR-II model contains the primary loop of the of the EBR-II system which consist of two centrifugal pumps (PSP1 and PSP2), a Low Pressure Plenum (LPP), a High Pressure Plenum (HPP), the active core, an Upper Plenum (UP) and an Intermediate Heat Exchanger (IHX). The two centrifugal pumps fulfill the task of liquid sodium circulation over the primary loop. The liquid sodium is drawn from the sodium tank by PSP1 and PSP2, which inject the coolant into the HPP and LPP. The HPP is connected to the active core to cool the EBR-II subassemblies, the control rods and inner reflector. The LPP directs the liquid sodium to the outer blanket. The hot coolant joins the "Z-pipe" from the UP and heads to the IHX. Boundary conditions impose the temperature of the argon and the sodium in the sodium tank on the inlet side. To simulate the heat sink that cools the primary loop by the intermediate loop, a simplified representation of the IHX on the intermediate loop side uses a flow and pressure boundary conditions. This simplified EBR-II model has been verified in terms of reactivity feedback, total core power, peak cladding temperature, and flow rate in primary pumps and the z-pipe. The level of consistency between the full-scale model and simplified EBR-II model is deemed sufficient to support the NAMAC proof of concept study presented in this paper. Three transients are simulated with the simplified GOTHIC model and compared against a validated SASSYS model [31]. The results of the simulations compare reasonably well. On the functional level this GOTHIC plant simulator operates in parallel with NAMAC to:

- Simulate the accident scenario for the case study;
- Generate virtual sensor data for plant state diagnosis;
- Accept control recommendations from NAMAC system.

The same system code (GOTHIC) is used to generate training/testing data for the development of DTs with different functions. We explicitly chose GOTHIC as both the plant simulator and data generation engine to isolate uncertainties (due to various data-driven models) during the testing phase.

The scheme for operation of data generation engine (RAVEN-GOTHIC interface) is described as follows. The EBR-II model is first modelled in steady state with GOTHIC to establish the normal operation regime of the plant. GOTHIC outputs a restart file representative of the steady state and a transient scenario is implemented through a GOTHIC-formatted text-based file called GOTHIC Command File (GCF). Perturbed variables (e.g. a pump speed change, a valve speed closure etc.) are flagged in the GCF to allow RAVEN sampling. A separate XML-based input file defines the RAVEN settings: number of samples ($N$), distribution type associated to the variables, variable bounds, and targeted outputs (i.e. potential features). The RAVEN execution replaces the variables drawn from the RAVEN source code into the GCF and runs successively the $N$ samples with the GOTHIC solver. At the end the $N$ GOTHIC executions, RAVEN parses the GOTHIC outputs (SGR files) and dumps the targeted parameters into a Comma separated file (CSV) or a HDF5 files. The CSV or HDF5 files are stored for DT training purposes. Figure 8 summarizes the workflow described above.

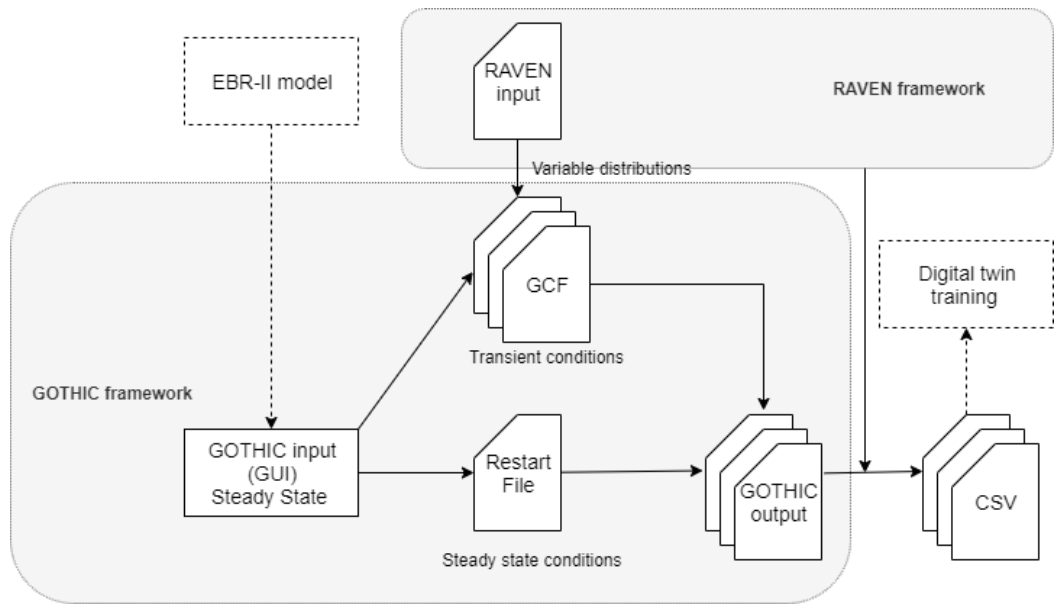

Figure 8: RAVEN/GOTHIC workflow

**2.4. NAMAC operational workflow**

Figure 9 shows the operational workflow of NAMAC system, the DT-D, and DT-P are highlighted. The workflow is developed based on the safety analysis literature [32], operating experiences, and control procedures of EBR-II. For the proof of concept study presented in this paper, the objective of the NAMAC system is to re-establish normal flow conditions during a single LOFA scenario. If that attempt fails, NAMAC should recommend to SCRAM the reactor. To achieve these objectives, a set of DTs, including diagnosis, prognosis, strategy inventory, and strategy assessment, are required, and they are connected according to the operational workflow. Among all DTs, the diagnosis and prognosis are critical in recovering full-flow conditions and preventing severe consequences, i.e., getting into the wrong region of the power/flow map and not promptly SCRAMming the reactor. The DT-D reads sensor data from the reactor or simulator to monitor unobservable state variables, and the objective is to restore complete information about the reactor. The obtained values are fed to (1) prognosis for predicting future transients (2) strategy inventory for determining available control actions (3) discrepancy checker for continuous monitoring. The DT-P reads reactor information from DT-D, together with available control actions, to predict the future transients of reactor states or consequences over a certain time range. The DT-P outputs are mainly used by strategy assessment for deciding the best strategies according to certain preference structures, which represent the preference of operator or risk management team to the economic and safety of a reactor.

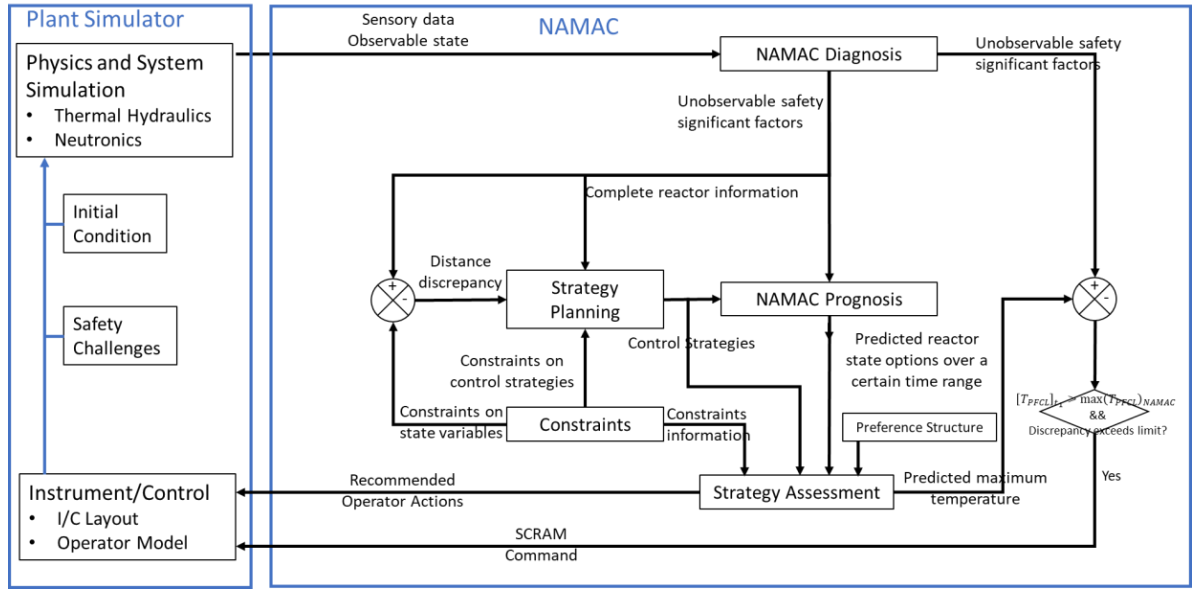

Figure 9: Operational workflow of a NAMAC system for controlling EBR-II during a single LOFA scenario

The operational workflow is designed based on the knowledge base. In this study, the emergency operating procedures and severe accident management guideline is adopted, and the goal is to build a safety case that prevents or mitigates from severe consequences. When a different advanced reactor system is investigated with different knowledge base, the operational workflow can vary from the present design. However, considering the nature of safety case as an argumentation toward safety goal, a similar reasoning process can be used. Moreover, it is argued that no matter how much *a priori* analysis gets done for advanced reactors, since there is no operational history, a system should still be doing active diagnosis, prognosis, strategy assessment, and discrepancy checking.

## 3. Demonstration of NAMAC Development and Assessment

This section presents a numerical demonstration of DTs and NAMAC system during the LOFA scenario. This demonstration is systematically illustrated based on objective, scope, DTs' implementation and assessment, GOTHIC-NAMAC interaction and operation procedure, success criteria, results, and implication of the case study.

### 3.1. Objective, Scope, and Assumption

This demonstration provides a proof of concept to demonstrate the capability of the NAMAC system as a control approach. The numerical demonstration is performed for a loss of flow due to a primary pump malfunction. In this case, NAMAC is expected to successfully diagnose the plant state, find available control actions, predict the consequence of each available action, and provide a plan of action that stabilizes the plant state.

The scope of the numerical demonstration is governed by the issue space of the case study. The issue space for this case study encompasses a partial loss of flow in Primary Sodium Pump#1

(PSP1) and is fully represented by time-dependent curves of rotational speed, $w_1(t)$, defined by Eq. 8,

$$w_1(t) = w_0 \left(1 - \frac{1 - (w_1)_{end}}{T_1} t\right), \quad t_{acc} + T_1 \geq t \geq t_{acc} \qquad \text{Eq. 8}$$

Here, $w_0$ represents the nominal pump speed (in rad/s), $T_1$ represents the ramping-down duration (in s) and $(w_1)_{end}$ represents the normalized pump #1 end speed. $t_{acc}$ marks the time when the accident transient begins. By varying the pump end speed, $(w_1)_{end}$, and pump ramp down duration, $T_1$, different profiles for pump ramp down can be achieved. According to the PRA analysis report [32], the ramping-down duration ($T_1$) is suggested to be less than 80 seconds for covering the most severe case. We select the accident scenario shown in Table 2 for demonstration of the NAMAC system for this case study. The scenario in Table 2 focuses on accident scenarios where PSP1 ramps down to 50% of its nominal speed in 50 seconds. This situation implies a pump coast-down rate of 1 rad/sec. The available control options to counter this situation are to accelerate the Primary Sodium Pump # 2 (PSP2) to normalized pump velocity $(w_2)_{end}$ for a fixed duration of time. Pump acceleration is triggered at time $t_{trip}$ which is defined by temperature thresholds called pump trip temperature ($T_{trip}$). $T_{trip}$ is based on the peak Fuel Centerline Temperature ($T_{FCL}$). The training data set to cover this scenario is generated by uniformly sampling $(w_2)_{end}$ and $T_{trip}$ with the sampling space defined by $[(w_2)_{min}, (w_2)_{max}]$ and $[(T_{trip})_{min}, (T_{trip})_{max}]$. $(w_2)_{end}$ and $T_{trip}$ are both uniformly sampled 32 times to generate 1024 episodes of 200 seconds transients. Each episode contains a 200-second transient from the same steady-state calculation that is loaded from a restart file (see Figure 8). Each sample output contains 2000 data points, and in total, there are about $2 \times 10^6$ data points in the training and testing databases.

Table 2: Description of accident scenario for demonstrations

| Initial condition | Accident scenario | Available control options | Constraints |
|---|---|---|---|
| Reactor is operating at nominal power $P_0$, and both pumps are operating at nominal rotational speed $w_0$ | Pump #1 malfunction: Pump #1 motor rotational velocity $w_1$ linearly coasts down to $(w_1)_{end}$ in $T_1$, $(w_1)_{end} = 50\%$, $T_1 = 50 sec$ (pump #1 coast down rate =1 rad/sec) | Increase PSP 2 rotational velocity $w_2$ towards $(w_2)_{end}$ for a fixed duration. Pump acceleration begins at time $t_{trip}$ defined by some threshold value called pump trip temperature ($T_{trip}$). | The reactor's peak fuel centerline temperature $T_{FCL}$ is picked as the safety significant factor. The $T_{FCL}$ is maintained at given set ranges $[(T_{FCL})_{low}, (T_{FCL})_{high}]$ |

Additional scenarios (see cases A and C in Table 7) with pump coastdown rate greater or less than 1 rad/sec are generated to evaluate the performance of the NAMAC system when it ventures outside the training domain. These scenario sets lie within the issue space but are not covered by the present training database.

For now, the measurement error in the sensor data is assumed to be negligible. The impact of component reliability is not considered in this case study. It is also assumed that the accident is known as soon as it is injected and that NAMAC makes recommendations as soon as the sensor information is fed (no time delays). This is achieved by pausing the GOTHIC plant simulator until recommendation are generated and injected. The objective is to minimize uncertainties for development and to have control over the sensitivity of the results to NAMAC's computational time. Furthermore, NAMAC needs a specific diagnosing time $t_D$ (in reactor-time scale) to collect information and make recommendations. Finally, the control strategies are assessed and ranked based only on the safety margins, while the reliability and lifetime of components are not explicitly considered. A range of control options for PSP2 is assigned based on the operating experiences of primary pumps in EBR-II [33].

### 3.2. Digital Twin Development

Based on the definition, DTs can be characterized by environment (DTE), prototype (DTP), and instance (DTI). Since DTs are treated as major components to the NAMAC control system, DTE reflects function and use case of DT, DTP suggests the modeling options, and DTI suggests the interface. Table 3 summarizes the implementation of the NAMAC DTs and discrepancy checker based on the functional, interfacing, and modeling requirements for this case study.

Table 3: Implementation of requirements for DTs in function (DTE), interface (DTI), and modeling (DTP)

|  | Function | Interface | | Modeling |
|---|---|---|---|---|
|  |  | Input | Output |  |
| DT-D | To monitor unobservable safety significant factor | Sensory Data ($X_D$): High pressure lower plenum temperature($T_{HPP}$), low-pressure lower plenum temperature ($T_{LPP}$), and upper plenum temperature ($T_{UP}$) | Unobservable safety significant factor: Fuel centerline temperature ($T_{FCL}$) | Static Feedforward Neural Networks<br>Training parameters ($P_D$)<br>• # of Layers: 3<br>• Neurons in each Layer: 20<br>• Max. Epochs: $10^6$<br>• Target mean square error: $10^{-2}$<br>Knowledge base $KB_D$: GOTHIC Simulation databases |
| DT-SI | To generate potential mitigating actions and corresponding control procedures | ▪ Constraints: pump#2 capacity $[(w_2)_{min}, (w_2)_{max}]$, Trip temperature $\left[(T_{trip})_{min}, (T_{trip})_{max}\right]$<br>▪ Safety significant factor ($X_{t0}$): Fuel Centerline Temperature ($T_{FCL}$) | Control procedures: trip points ($T_{trip}$) and control inputs ($w_2)_{end}$ | Uniform sampling |
| DT-P | To predict safety significant reactor state variables over a time range | ▪ Safety significant factors ($X_{t0}$): initial value of fuel centerline temperature and its time derivative from diagnosis $\left[T_{PFCL}, \overline{\frac{dT_{PFCL}}{dt}}\right]_{t_0}$<br>▪ Control procedures: trip points ($T_{trip}$) and control inputs (($w_2)_{end}$) from strategy planning, i.e., $\left[(w_2)_{end}, T_{trip}\right]$<br>▪ Constraint on fuel centerline temperature, $(T_{FCL})_{High}$ | A class of global factors (prediction of safety significant reactor state variables) over a time range (200 s transient) | Static Feedforward Neural Networks<br>Training parameters ($P_p$):<br>• # of layers: 3<br>• Neuron in each layer: 20<br>• Max. Epochs: $10^6$<br>• Target mean square error: $10^{-3}$<br>Knowledge base $KB_P$: GOTHIC Simulation databases |
| DT-SA | To rank control procedures and make recommendations | ▪ Preference structure (safety limit for the predicted state variable)<br>▪ Control procedures (trip points and control inputs from strategy planning)<br>▪ Predicted state variables (from prognosis)<br>▪ Constraints | ▪ Recommended operator actions<br>▪ Recommended global factor and short-term state transients | ▪ Simple decision analysis using limit surface which identifies the set of control inputs that causes transition of label from safe to unsafe<br>▪ Safety margin is defined based on the preference structure to rank control strategies |

The performance of data-driven DTs (i.e., DT-D and DT-P) is dominated by the error between predicted and real values. Therefore, an accuracy criterion is specified by users' knowledge. To ensure the fairness of this assignment, the value of the criteria has been reviewed by members with backgrounds in both data science and nuclear STH. A distance error metric, Root Mean Square Error (RMSE) between predicted and real values is used to quantify the accuracy of DTs, and it is required that RMSEs are less than 5% of DT outputs or Quantities of Interest (QoIs). The accuracy criteria are not fixed, and their assignments need to be formalized by a decision-theoretic framework.

$$RMSE = \sqrt{\frac{1}{N}\sum_{i=1}^{N}(QoI_{preds} - QoI_{obs})^2} \qquad \text{Eq. 9}$$

A three-layer feedforward neural network is constructed by MATLAB, and Bayesian regularization backpropagation is used as the learning algorithm for both diagnosis and prognosis DTs. For DT-D model, the peak fuel centerline temperature $T_{PFCL}$ is selected as the unobservable state variable that is safety significant, and it is evaluated based on sodium temperature measured at three locations inside the core: UP sodium temperature $T_{UP}$, HPP sodium temperature $T_{HPP}$, and LPP sodium temperature $T_{LPP}$. As a result, variables and values for sets of inputs, training parameters, and knowledge bases as in Eq. 1 can be selected (see DT-D in Table 3 for details). The DT-SI uniformly samples the control options based on the evaluated value of unobservable safety significant factor obtained from diagnosis (see DT-SA in Table 3). The DT-P predicts the transient of safety significant reactor state variable and consequences over a specified duration of time. The maximum $T_{PFCL}$ in 200 sec after the steady state restart time is selected as the QoI, and it is predicted based on a set of parameters and variables as in Eq. 2 (see DT-P in Table 3 for details). The DT-SA is performed by finding the limit surface where control inputs cause transition from safe to unsafe consequences (see DT-SA in Table 3 for details).

Impact of component reliability on the strategy is not included in this initial demonstration study. It should be noted that the diagnosis and prognosis models presented in Table 3 are based on preliminary implementations (baseline option) and have been individually trained and tested for the accident scenario specified in Table 2. However, the performance of these models in the integral NAMAC system needs to be determined.

In Table 3, $[(w_2)_{end}, T_{trip}]$ is the control action set $A$; $(w_2)_{end}$ is the final and steady rotational speed of the primary pump; $T_{trip}$ is the temperature point that triggers the injection of pump actions $w_{end}$. $\left[T_{PFCL}, \overline{\frac{dT_{PFCL}}{dt}}\right]_{t_0}$ is the information set $X$ at time point $t_0$; $[T_{PFCL}]_{t_0}$ represents the peak fuel centerline temperature at time $t_0$, while $\left[\overline{dT_{PFCL}/dt}\right]_{t_0}$ represents the time gradient of peak fuel centerline temperature at time $t_0$; the bar notation suggests that this number is approximated numerically by a finite difference scheme as shown in Eq. 10:

$$\frac{\overline{dT_{PFCL}}}{dt} = \frac{([T_{PFCL}]_{t_0} - [T_{PFCL}]_{t_0-\Delta t})}{\Delta t} \qquad \text{Eq. 10}$$

Here, $\Delta t$ is the time range for numerically approximating the time gradient of peak fuel centerline temperature. To account for the effect of this number, three time ranges are used at the same time for redundancy. It is suggested that the feature selections have been embodied inside the regularization methods during the neural network training. As a result, the insignificant variables will be automatically reduced and reflected in the weights and biases within the networks. Figure 10 compares the predicted outputs by DT-D and DT-P against real values from databases. Most points fall onto the 45-degree line, implying that good accuracy is achieved for both DT implementations.

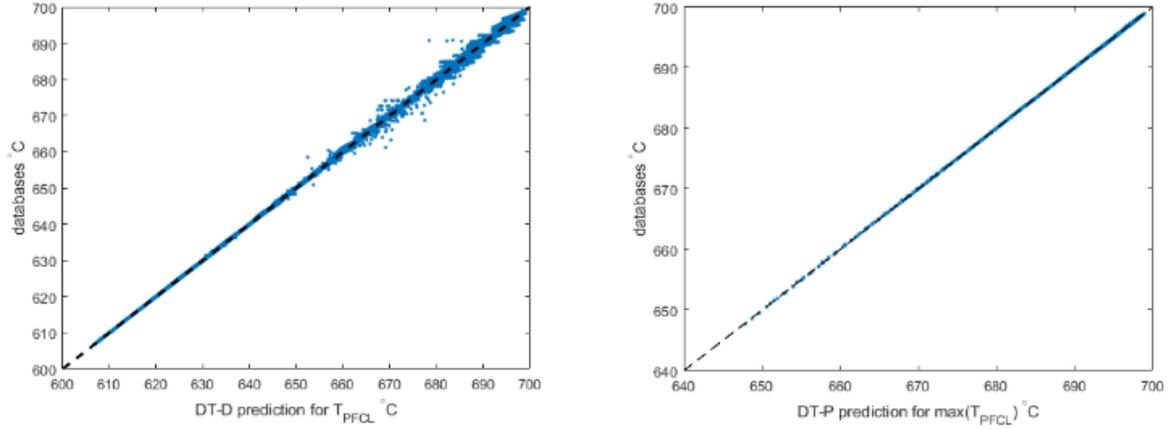

Figure 10: Comparison of predicted output against real values from databases for DT-D (left) and DT-P (right) models. Dash lines are diagonal lines

The RMSEs of DT-D and DT-P in the training process are 1.42℃ and 0.54℃ respectively. Both errors satisfy the acceptance criteria that the RMSE is less than 5% of QoIs ($[-30, 30]$℃). The DT-SI generates control strategies based on the DT-D outputs, safety and control limits. In the current implementation, control actions are generated by uniform distributions for trip temperature $T_{trip}$ and PSP2 end speed $(w_2)_{end}$. The DT-SA ranks the control procedure and makes recommendations for operator actions. In the current implementation, strategy assessment is based on the limit surface. The limit surface identifies the set of control inputs that causes transition of the SSF from safe to unsafe region. By defining the safety margin for each available control action $i$ as Eq. 11, the acceptable region and the optimal actions can be identified.

$$S_i = (T_{PFCL})_{lim} - [max(T_{PFCL})]_i. \qquad \text{Eq. 11}$$

where $(T_{PFCL})_{lim}$ is the safety limit for the Peak Fuel Centerline Temperature (PFCL). In this study, the number is assigned based on user's knowledge. $[max(T_{PFCL})]_i$ is the maximum PFCL predicted by the DT-P model for control action $i$. Figure 11 (a) depicts the safety region with positive safety margins $S_i$, which represents a set of control options, including trip temperature $T_{trip}$ and PSP2 end speed $(w_2)_{end}$ that will keep maximum PFCL $max(T_{PFCL})$ below the safety limit $(T_{PFCL})_{lim}$. Figure 11 (b) shows the variation of that region due to different selections of safety limits. Although the region is changed due to safety limits, the optimal control actions with maximum margin remain the same.

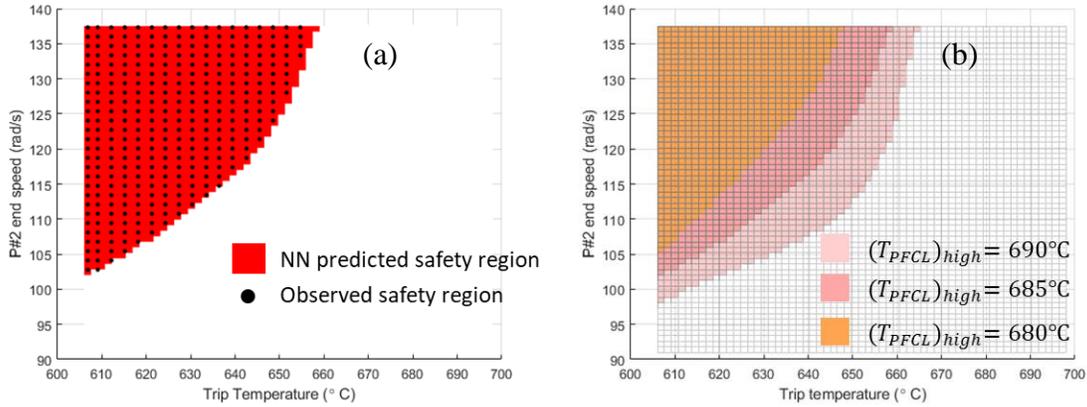

Figure 11: Plot of safety regions (a) predicted by NAMAC versus observations from the GOTHIC plant simulator with $(T_{PFCL})_{lim}=685°C$. (b) with three different $(T_{PFCL})_{lim}$.

### 3.3. Digital Twin Assessment

Since DTs are integrated into the NAMAC control system, and they potentially have significant impacts on the reactor safety, it is critical to assess the accuracy of DT implementations not only in the training process, but also during testing and applications. In this study, a list of sources of uncertainty is prepared based on classifications by M. Kläs *et al.* [34] Meanwhile, the error assessment works are inspired by Verification, Validation and Uncertainty Quantification (VVUQ) guidelines for scientific computing [24].

Table 4 shows the list of sources of uncertainty and their relative impacts on the DTs' accuracy based on the qualitative assessment results. It is stressed that the current assessments are at scoping stage, where the acceptance criteria, sources of uncertainty, samples and distributions of parameters, even the criteria for deciding relative impacts, are made by user knowledge and subject to refinement. For the current implementation of DT for diagnosis (DT-D) and selection of training and testing databases, it is found that in both training and testing cases, the accuracy of DT-D is highly correlated to input time ranges, sensor failures, final training loss, and database coverage. For the current implementation of DT-P, it is found that in training case, the DT-P accuracy is highly correlated with the final training loss. However, that correlation is not observed in the testing case. Moreover, the DT-P accuracy is found to be highly correlated with the database coverage when DT-P is applied to a different testing case. In consequence, considering the complexity of transients and difficulty in making predictions for future states, it is suggested that advanced machine learning algorithms should be adopted, especially for prognosis, such that the complexity of networks is comparable to data. Meanwhile, since the implementation of DTs is highly affected by multiple sources of uncertainty, it is also suggested that a formalized DT development and assessment process should be developed. The objective is to implement DTs in a transparent, consistent, and improvable manner, where refinements on DTs can be made based on the previous development and assessment results. More details in results are discussed in Appendix A.

Table 4: A list of sources of uncertainty, formulations, and findings for data-driven DT models in the preliminary quantitative assessment

| Category | Sources of uncertainty | Parameter range | Findings |
|---|---|---|---|
| **Digital Twin for Diagnosis** | | | |
| Input variables | Input time ranges $t_D$ | [0,200] | Errors are negligible when DT-D is monitoring $T_{PFCL}$ within a time range of 35s after $t_0$ with given networks and training algorithms |
| Input variables | Sensor failure | Inputs with N/A values | Errors are more sensitive to sensor failures at high-pressure plenum with given networks and training algorithms |
| Training parameters | Final training loss $\varepsilon_D$ | $[10^{-3}, 10]$ | Errors are strongly affected by accuracy requirements with given networks and training algorithms |
| Knowledge base | Database coverage | Mutual information between training and testing data distributions | Errors are strongly affected by coverage conditions based on the current coverage metrics |
| **Digital Twin for Prognosis** | | | |
| Training parameters | Final training loss $\varepsilon_D$ | $[10^{-3}, 10]$ | Errors are only slightly affected by accuracy requirements with given networks and training algorithms |
| Knowledge base | Database coverage | Interpolated and extrapolated conditions | Errors are strongly affected by coverage conditions based on the current coverage metrics |
| **Digital Twin for Strategy Inventory** | | | |
| Input variables | Error propagated from DT-D | | Error in DT-D strongly affects the selection of control actions |
| Knowledge base | Control parameter space coverage | | Sampling grid resolution strongly affects the efficacy of the control parameter space coverage |
| Input variables | Error propagated from DT-D | | Error in DT-D strongly affects the selection of control actions |
| **Digital Twin for Strategy Assessment** | | | |
| Input | Error propagated from DT-P | | Error in DT-P and DT-SI leads to error in recommended control action |
| Input | Error propagated from DT-SI | | |
| Knowledge base | Selection of safety limits | | Selection of safety limits does not affect the recommendation for optimal control actions |

### 3.4. NAMAC-GOTHIC Coupled System Development

A coupled-system interface has been created for explicit coupling of NAMAC and GOTHIC plant simulator model. The timeline of different events and sequence of processes for NAMAC-GOTHIC interaction are described in the following steps (see Figure 12 and Figure 13 for illustration):

i. A waiting period of $t_w$ is assumed during which the plant simulator operates under steady state. The steady state conditions are provided by the .SDM file from GOTHIC. At time $t_{acc} = t_{rmcd} - t_D$, fault is injected by ramping down the rotational speed of PSP1 or reducing the torque to a specific value. Here, $t_D$ is the NAMAC time range for diagnosis. When time $t_{rmcd} = t_{acc} + t_D$ is reached (step ① in Figure 12), the GOTHIC plant simulator is paused, and it resumes operation when new actions are injected by NAMAC.

ii. GOTHIC output files (.SGR file containing sensor data ) are processed and injected into the NAMAC diagnosis and discrepancy checker for reactor-state analysis and control (step ② in Figure 12).

iii. NAMAC diagnoses the plant state and makes recommendations based on the inherent DTs and sensory inputs (step ③ in Figure 12). NAMAC diagnosis monitors the plant state (fuel centerline temperature) based on the observable sensor data (see Table 3 for details). The strategy inventory provides the control procedures based on the constraints and safety significant factor. The prognosis provides global factors (i.e., maximum fuel centerline temperature) during a predefined time range (around 200 seconds). The strategy assessment ranks the control action based on the safety margin and makes recommendations for operator actions. If no available action can be found that satisfies the safety criteria, a SCRAM signal is generated to cease the plant operation.

iv. The discrepancy checker gets into action after the control action based on the NAMAC recommendation(s) is injected (i.e., at time $t_1 > t_{rcmd}$). It employs the diagnosis model to monitor the peak fuel centerline temperature and compares it with predicted peak fuel centerline temperature obtained from NAMAC prognosis. If the discrepancy between these two values exceeds a predefined limit ($X_{lim}$), a SCRAM signal is injected. However, if the discrepancy is less than $X_{lim}$, then no action is taken. In this way, discrepancy checker adds another layer of safety to anticipate unexpected NAMAC predictions triggered by high uncertainties.

$t_{acc}$ is the time for injecting pump malfunctions; $t_{rmcd}$ is the time for making and executing recommendations by NAMAC; $t_D$ is the time difference between $t_{acc}$ and $t_{rmcd}$, where NAMAC collects information from sensors. The GOTHIC plant simulator is paused at $t_{rmcd}$ such that the computational time of NAMAC is neglected.

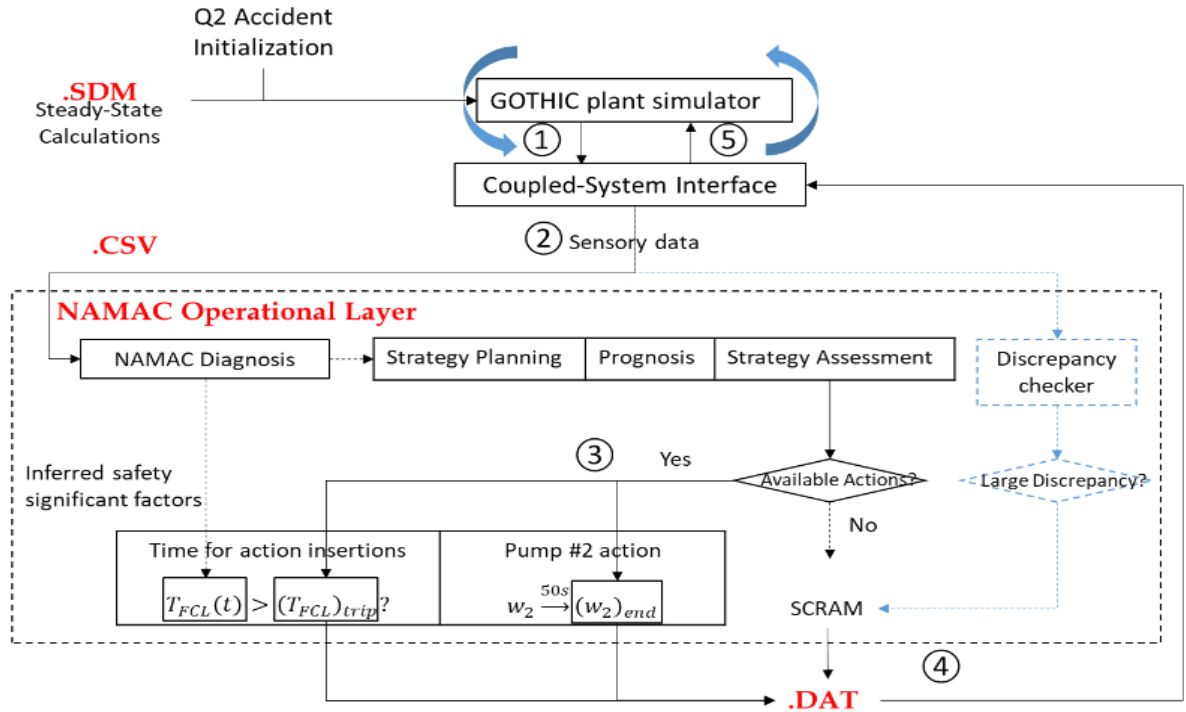

Figure 12: NAMAC operation with GOTHIC plant simulator

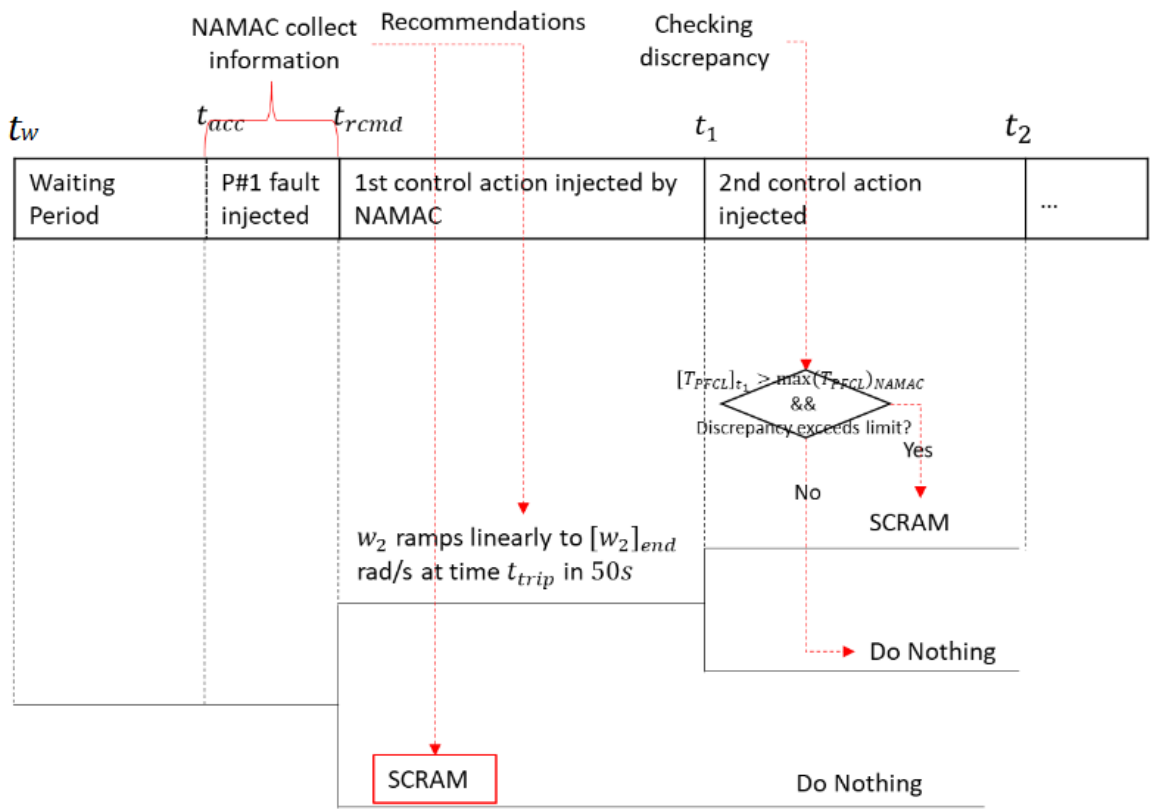

Figure 13: Timeline of events during NAMAC-GOTHIC interaction

### 3.5. NAMAC-GOTHIC Coupled System Assessment

The success criteria for the NAMAC "proof of concept" demonstration are governed by the capability of the NAMAC system to correctly diagnose the plant state, find available control actions, predict transient and consequence of each action, and make recommendations based on a user-defined preference structure. As the data-driven methods for different DTs (specifically, diagnosis and prognosis in the current implementation) exhibit inherent modeling uncertainties, some errors in the prediction are expected. Therefore, a graded approach (see Table 5 for details) is adopted to assess the accuracy of NAMAC and to define the success criteria. To quantify the uncertainty of NAMAC recommendations, predictions of the maximum peak fuel centerline temperature $T_{PFCL}$ during the entire transient are made and compared against simulations by the GOTHIC plant simulator. This prediction is used to justify if the recommendation is reasonable. Comparisons are made against the observed values for assessing the performance of NAMAC system and bounds of the target uncertainty for diagnosis is given by:

$$\varepsilon_D \leq 5\% QoI \qquad \text{Eq. 11}$$

Table 5: Success criteria for numerical demonstration

| Grade | Description |
|---|---|
| Level 0 | Failure - NAMAC fails to diagnose the correct plant state |
| Level 1 | Partial success- NAMAC diagnoses the plant state within the bounds of target uncertainties for diagnosis but fails to provide recommendations that would stabilize the reactor state. Therefore, reactor is SCRAMed |
| Level 2 | Success- NAMAC diagnoses the plant state within the bounds of target uncertainties for diagnosis and provides recommendations that stabilizes the reactor state. |

Initial conditions are loaded from the .SDM files, and the plant simulator is restarted from the established steady state. The waiting period with steady state condition is 10,000 sec and under this condition, the pumps operate at their nominal speeds, sodium and fuel temperature profiles are established across the primary loop of the EBR-II system. The accident scenario is injected into the plant simulator at time $t_{acc}$=10,010 sec. The accident scenario involves ramping down of PSP1 to 50% for a duration of 50 sec (i.e., the pump coast down rate =1 rad/sec). At time $t_{rcmd} = 10,020$, the plant simulator is paused and a data file (containing sensor data) is generated and passed to NAMAC for making recommendations. The output data file (simp-data-T.dat) is fed into NAMAC for:

- Diagnosing the unobservable plant states (peak fuel centerline temperature);
- Generating control actions (sampled PSP2 speed and trip temperature);
- Predicting the consequences (maximum peak fuel centerline temperature for 150s transient);
- Assessing control actions (safety margins);
- Making recommendations (maximum safety margins).

Figure 14 shows the error in the predicted fuel centerline temperature obtained from diagnosis. Figure 15 shows the contour plot of safety margins against the available control actions, i.e., trip temperature and PSP2 speed at time $t_{rcmd} = 10,020$ sec. Based on the inherent DTs and NAMAC systems, NAMAC recommends an action for PSP2 or for reactor SCRAM:

- At time $t_{rcmd} = 10,020$s, 10 sec after the accident being injected, NAMAC recommends to ramp up the PSP2 speed to maximum (150%) immediately (at $t_{rcmd}$).
- These pump conditions and reactor settings correspond to the maximum safety margin (+20.9°C).
- The maximum diagnosis error is 13°C (acceptance criteria $\varepsilon_D \leq 5\% QoI$).

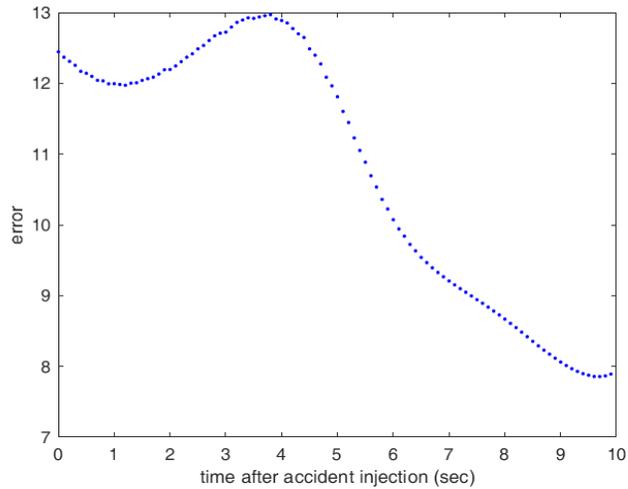

Figure 14: Diagnosis error (°C) in the time range $[t_{acc}, t_{rcmd}]$

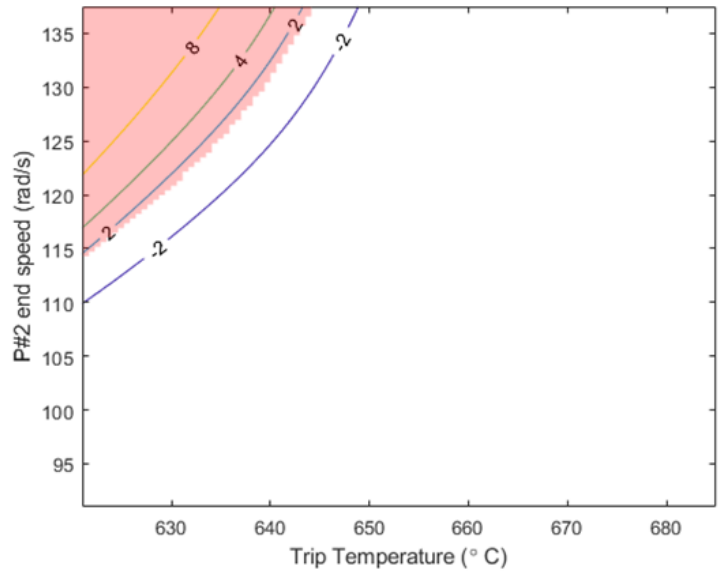

Figure 15: Contour plot of safety margins against the available control actions: trip temperature and PSP2 speed, where shaded area represents the acceptable region for control actions with positive safety margins

If the operator decides to ignore the recommendation and do nothing (see Figure 16), the temperature will exceed the safety criteria (685°C) and reach 698.8°C at $t = 10,063.5$s. If the operator decides to follow NAMAC recommendations (see Figure 17), the maximum temperature is 655.5°C, which is below the safety criteria. The predicted maximum temperature by NAMAC is 664.1°C. These results indicate that numerical demonstration with fully coupled system of NAMAC and GOTHIC plant simulator is completed with level 2 success criteria (see Table 5) when the accident scenario is covered by the training data set.

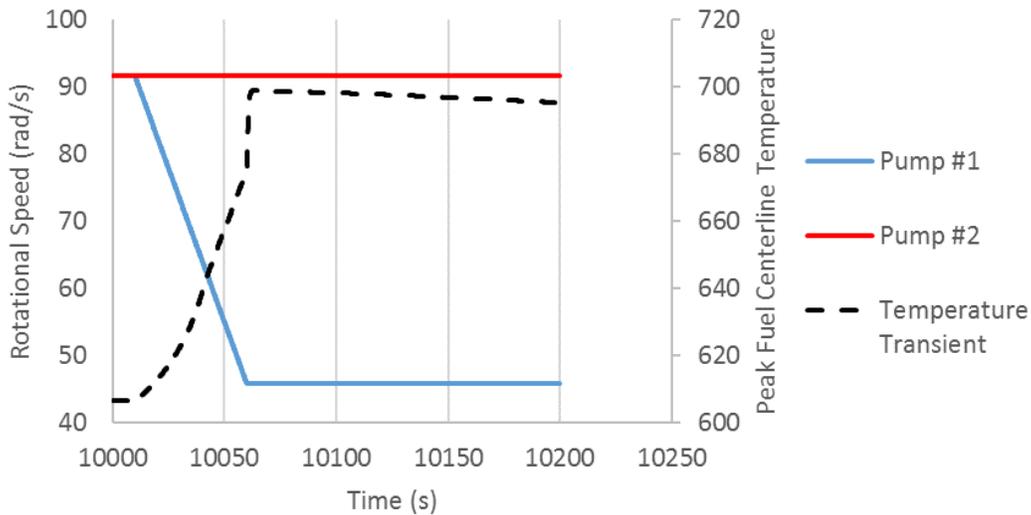

Figure 16: Transient of $T_{PFCL}$, no PSP2 action is taken after the injection of PSP1 malfunction

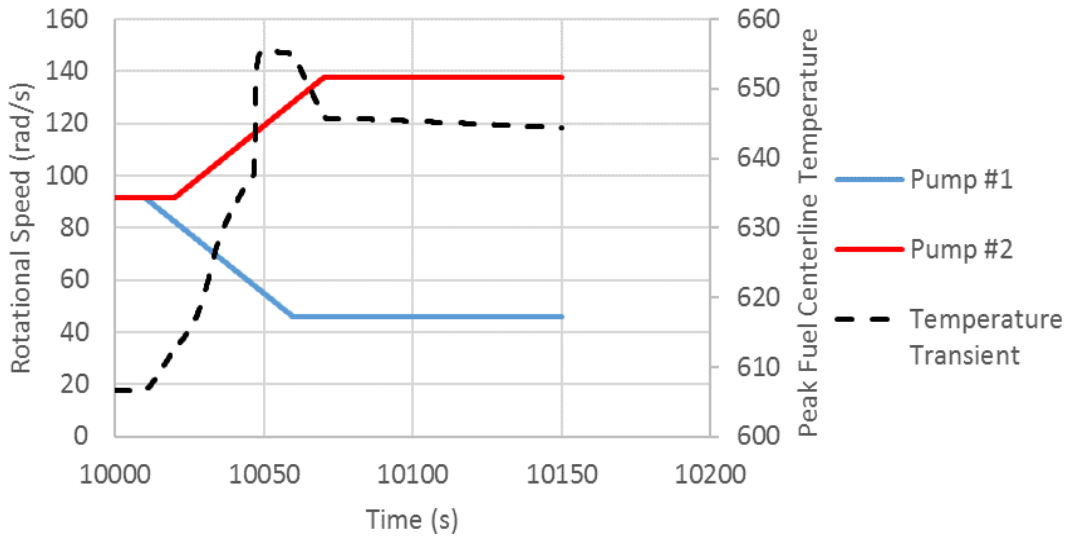

Figure 17: Transient of $T_{PFCL}$, following NAMAC recommendation to increase PSP2 speed.

Although NAMAC provides acceptable recommendations for the specified accident scenario, it is important to analyze the scenario adaptability when it encounters scenarios that are not covered by the training domain. Therefore, the NAMAC system is tested against additional scenarios with different pump coast down rates $[(w_1)_{end}, T_1]$ and NAMAC diagnosis time $t_D = [1s, 10s]$. These scenarios are categorized into three sets, A, B, and C, based on different pump coast down rates as shown in Table 7. Scenarios set A contains severe transients with larger pump coast down rates than those in training databases. Case C constitutes moderate transients with slower pump coast down. Scenarios set A and C are not covered by the training data. The coast-down rate of scenario set B is the same as that in the training data. However, the pump end speed $(w_1)_{end}$ and coast down duration $T_1$ are different. Based on the predicted temperature by NAMAC and the real temperature from the plant simulator, errors of the NAMAC system can be determined: if the maximum temperature is lower than a limit, the scenario is classified as safe, otherwise, the scenario is unsafe. Therefore, it is possible that the predicted and real states are inconsistent when the system errors of NAMAC are too large.

A confusion matrix [35] is defined to visualize and quantify the accuracy of NAMAC system in classifying the predicted consequence of control actions. Table 6 shows a table of confusion that defines four conditions: False Positive (FP), False Negative (FN), True Positive (TP), and True Negative (TN). Each row of the matrix represents the instances in a predicted class while each column represents the instances in an actual class. These four conditions are defined as follows:

- False Positive (FP): True state is unsafe, but NAMAC prediction is safe;
- False Negative (FN): The true state is safe, but NAMAC prediction is unsafe;
- True Positive (TP): True state is safe, and NAMAC prediction is also safe;
- True negative (TN): True state is unsafe, and NAMAC prediction is also unsafe.

Among the four categories, TP and TN suggest correct classifications by NAMAC recommendations. FP and FN represent confusing classifications: FP tends to have the most significant impacts as NAMAC underpredicts the consequences; FN indicates that NAMAC is making conservative recommendations, and the severity of consequences are overpredicted. To have quantified results in addition to visualizations, True Positive Rate (TPR), False Negative Rate (FNR), False Positive Rate (FPR), and True Negative Rate (TNR) are defined to characterize the accuracy of the classification systems. Both the FNR and FPR represent the errors of NAMAC outputs, and higher values suggest larger errors. Since FP conditions could cause severe consequences, the acceptance criteria for FPR are expected to be lower than FNR.

Table 6: Definition of confusion matrix with typical confusion rates

|  |  | **True Condition** | |
|---|---|---|---|
|  |  | Condition Positive | Condition Negative |
| **Predicted Condition** | Predicted Condition Positive | True Positive | False Positive |
|  | Predictive Condition Negative | False Negative | True Negative |
|  |  | True Positive Rate = $\frac{\sum \text{True Positive}}{\sum \text{Condition Positive}}$ | False Positive Rate = $\frac{\sum \text{False Positive}}{\sum \text{Condition Negative}}$ |
|  |  | False Negative Rate = $\frac{\sum \text{False Negative}}{\sum \text{Condition Positive}}$ | True Negative Rate = $\frac{\sum \text{True Negative}}{\sum \text{Condition Negative}}$ |

Since the classification is made based on outputs from a sequence of DTs, results by confusion matrix represent the aggregation and propagation of DTs errors based on the operational workflow. The results of evaluation for the NAMAC system for cases A, B, and C are shown in Table 7. For scenario B, which is covered by the training data base, 100% TPR is observed. For moderate transients with slow pump coast down rates (i.e., case C), higher TPR is achieved, together with a small FNR. However, significant FPR is observed for case A, which contains severe accident scenarios outside the training data base. These results are expected since the uncertainty of data-driven methods grow drastically when they are making predictions outside the training domain. It should be noted that the Discrepancy Checker was excluded from the NAMAC system for the evaluation presented in Table 7. In the occurrence of a false positive scenario, the Discrepancy Checker detects an anomaly and recommends to immediately SCRAM the reactor.

Table 7: NAMAC performance for different accident scenarios

| ID | Pump#1 coast down rate (m) in rad/s | Diagnosis time | FPR | TPR | FNR | TNR |
|---|---|---|---|---|---|---|
| A | $10 > m > 1$ | [1s,10s] | 60% | 40% | 0 | 0 |
| B | $m = 1$ | [1s,10s] | 0 | 100% | 0 | 0 |
| C | $0.1 < m < 1$ | [1s,10s] | 10% | 90% | 10% | 90% |

## 4. Concluding remarks

This work illustrates the process of development, implementation and assessment of a Nearly Autonomous Management and Control (NAMAC) system to provide recommendations to the operator for maintaining the safety and/or performance of the reactor. The development process of the NAMAC system is illustrated by a three-layer hierarchical workflow that consists of knowledge base, Digital Twin (DT) development layer, and NAMAC operational layer. DTs play a critical role in NAMAC and are described as knowledge acquisition system to support different autonomous control functions. Based on the knowledge base, a set of DTs are trained to determine the plant state, predict behavior of physical components or systems for candidate control actions, and, based on that prognosis, rank available control options. The trained DTs are assembled according to the NAMAC operational workflow to support the decision making process for selecting the optimal choices during an accident scenario.

To demonstrate the capability of the NAMAC system, a case study is presented, where a baseline NAMAC is implemented for operating a simulator of the Experimental Breeder Reactor II (EBR-II) during a partial Loss Of Flow Accident scenario. A simplified five channel EBR-II model in GOTHIC is used as the plant simulator that generates data for knowledge base and tests NAMAC performance with the NAMAC-GOTHIC coupled system. The database for development of different data-driven DTs is obtained by sampling the scenario and control parameters using the RAVEN-GOTHIC data generation engine. Since the data-driven methods behave as a "black box", a list of sources of uncertainty, including input variables, training parameters, and knowledge base, is prepared to assess their relative impacts on DT errors. Based on the current implementation of Digital Twin for Diagnosis (DT-D), Digital Twin for Prognosis (DT-P), Digital Twin for Strategy Inventory (DT-SI), and Digital Twin for Strategy Assessment (DT-SA), Table 8 summarizes the relative impact of each source of uncertainty. It is stressed that the current assessments are at scoping stage, where the acceptance criteria, sources of uncertainty, samples and distributions of parameters, even the criteria for deciding relative impacts, are made by user knowledge and subject to refinement.

Table 8: A list of sources of uncertainty for each data-driven DT models and their relative impacts on the DT's performance

| Category | Sources of uncertainty | Relative Impacts |
|---|---|---|
| **DT-Ds** | | |
| Input variables | Input time ranges $t_D$ | Negligible |
| | Sensor failure | High to sensor failures at high-pressure plenum |
| Training parameters | Final training loss $\varepsilon_D$ | High |
| Knowledge base | Database coverage | High |
| **DT-P** | | |
| Training parameters | Final training loss $\varepsilon_D$ | Low |
| Knowledge base | Database coverage | High |
| **DT-SI** | | |
| Input variables | Error propagated from diagnosis | High |
| Knowledge base | Control parameter space coverage | High |
| Input variables | Error propagated from DT-D | High |
| **DT-SA** | | |
| Input | Error propagated from DT-P | High |
| | Error propagated from DT-SI | High |
| Knowledge base | Selection of safety limits | No |

For the current implementation of DT-D and selection of training/testing databases, it is found that in both training and testing cases, the accuracy of DT-D is highly correlated to all listed sources of uncertainty. For the current implementation of DT-P, it is found that in training case, the DT-P accuracy is highly correlated with the final training loss. However, that correlation is not observed in the testing case. Moreover, the DT-P accuracy is found to be highly correlated with the database coverage when DT-P is applied to a different testing case. As a result, it is suggested that advanced machine learning algorithms should be adopted, especially for prognosis, such that the complexity of networks is comparable to that in data. Meanwhile, since the implementation of DTs is highly affected by multiple sources of uncertainty, it is also suggested that a formalized DT development and assessment process should be developed. The objective is to implement DTs in a transparent, consistent, and improvable manner, where refinements on DTs can be made based on the previous development and assessment results.

After the training and testing, the DTs are assembled to form the integral NAMAC system. This baseline NAMAC system is coupled with the GOTHIC plant simulator for its demonstration and assessment. Three sets of accident scenarios are designed, and a confusion matrix is created to determine the misclassification ratios of NAMAC. It is found that within the training databases, NAMAC can make reasonable recommendations with zero FNR and FPR such that the peak fuel centerline temperature can also be maintained below the limit. However, when the scenario is beyond the training cases, the FNR and FPR increase, especially when the scenarios are more severe. Therefore, a discrepancy checker is suggested to detect unexpected reactor states and to alert operators for safety-minded actions. Moreover, control strategies in this NAMAC system are assessed and ranked based only on safety margins. The reliability or lifetime of primary sodium pumps, which adversely affect the pump performance, will be considered in the future.

## Abbreviation

| | |
|---|---|
| AI | Artificial Intelligence |
| DT | Digital Twin |
| DTE | Digital Twin Environment |
| DTP | Digital Twin Prototype |
| DTI | Digital Twin Instance |
| DT-D | Digital Twin for Diagnosis |
| DT-P | Digital Twin for Prognosis |
| DT-SA | Digital Twin for Strategy Assessment |
| DT-SI | Digital Twin for Strategy Inventory |
| EBR-II | Experimental Breeder Reactor II |
| FCL | Fuel Centerline Temperature |
| FDD | Fault Detection and Diagnosis |
| FHF | Function-based Hierarchical Framework |
| FN | False Negative |
| FNN | Feed Forward Network |
| FNR | False Negative Rate |
| FP | False Positive |
| FPR | False Positive Rate |
| HPP | High-Pressure Plenum |
| IHX | Intermediate Heat Exchanger |
| LOFA | Loss of Flow Accident |
| LPP | Low-Pressure Plenum |
| NAMAC | Nearly Autonomous Management and Control |
| NPP | Nuclear Power plant |
| PFCL | Peak Fuel Centerline Temperature |
| PSP | Primary Sodium Pump |
| PRA | Probabilistic Risk Assessment |
| QoI | Quantity of Interest |
| RMSE | Root Mean Square error |
| SCS | Supervisory Control System |
| SSF | Safety Significant Factor |
| TN | True Negative |
| TNR | True Negative Rate |
| TP | True Positive |
| TPR | True Positive Rate |
| UP | Upper Plenum |

**Acknowledgement**

This work is performed with the support of ARPA-E MEITNER program under the project entitled:" Development of a Nearly Autonomous Management and Control System for Advanced Reactors" and using a GOTHIC license provided by Zachry Nuclear Engineering, Inc.

GOTHIC incorporates technology developed for the electric power industry under the sponsorship of EPRI, the Electric Power Research Institute.

The authors would also like to acknowledge the comments and suggestions from Dr. Cristian Rabiti and Dr. Paolo Balestra of Idaho National Laboratory, Dr. Min Chi of North Carolina State University, Mr. John Link of Zachry Nuclear Engineering, Dr. Botros Hanna and Dr. Son Tran of New Mexico State University, Dr. David Pointer, Dr. Sacit Cetiner and Dr. Birdy Phathanapirom of Oak Ridge National Laboratory, and Dr. Carol Smidts of Ohio State University.

# Appendix A. Digital Twin Assessment Results

This Appendix discusses detailed results from digital twin (DT) assessment with respect to each source of uncertainty.

## A.1. Coverage Assessment for Digital Twin for Diagnosis

In this study, the coverage is firstly defined based on the characterization of accident scenarios and control procedures – e.g. end speeds or ramping-down speed for primary sodium pump (PSP) #1 and #2, trip temperature for action injection – for the training and testing scenarios. To evaluate the impact of coverage, testing scenarios with different characterizations than the training scenarios are used to assess the performance of each DT. Table 9 shows a list of training and testing case studies with different coverage conditions according to the characterization for accident scenario. Based on the distribution of testing and training data points, the coverage condition can be further classified into two classes: interpolation and extrapolation. Interpolation is defined as the conditions where no new data point exist in the testing database or the new data points in the testing database is within the range of a discrete set of known data points in the training domain. The extrapolation is defined as the conditions where new data points in the testing database is outside the range of training data points. Eq. 8 defines the issue space for this case study encompasses a partial loss of flow in Primary Sodium Pump#1 (PSP1) and is fully represented by time-dependent curves of rotational speed, $w_1(t)$. Again, $w_0$ represents the nominal pump speed (in rad/s), $T_1$ represents the ramping-down duration (in s) and $(w_1)_{end}$ represents the normalized pump #1 end speed. $t_{acc}$ marks the time when the accident transient begins. By varying the pump end speed, $(w_1)_{end}$, and pump ramp down duration, $T_1$, different profiles for pump ramp down can be achieved.

Table 9: Comparisons of DT-D errors with different coverage conditions

|  | $T_1$ | $(w_1)_{end}$ | Coverage condition | DT-D Errors |
|---|---|---|---|---|
| Training 1 | 21.02 | 51.6-100 | Global Extrapolation as $(\omega_2)_{end}$ in testing 1 is outside the range of training 1 | 0.01 |
| Testing 1 | 21.02 | 3.2 |  | 43.3 |
| Training 2 | 50 | 9.7-58.1 | Global Interpolation as $(\omega_1)_{end}$ in testing 2 is inside the range of training 2 | 0.10 |
| Testing 2 | 50 | 38.7 |  | 0.08 |

It is found that the DT-D performance depends heavily on the coverage conditions: when the characterization of testing scenarios locate outside the training scenarios, the DT-D error is grow larger than the training error; when the characterization of testing scenarios locate inside but not completely the same as the training scenarios, the DT-D errors are of similar order of accuracy to the training errors.

In addition to the global characterization, the coverage is defined by the data distribution and further quantified as the mutual information between the probability distribution function (PDF) of training and testing data points. In this study, a database is selected from the whole data repository, and all data points are approximated by the kernel density estimation as in Eq. 12:

$$PDF(P_i) = \frac{1}{n}\sum_{i=1}^{n} K_H(x - y_i) \quad \text{Eq. 12}$$

where $x$ represent the multi-dimensional random vectors with density function $P_i$; $y_i$ represents a random sample drawn from the database; $n$ represents the total number of data points in the selected database. Meanwhile, the mutual information is calculated with the symmetric Kullback-Leibler divergence (K-L divergence) as in

$$D_{KL}(P,D) = \sum_i P(i)\log\left(\frac{P(i)}{D(i)}\right) + D(i)\log\left(\frac{D(i)}{P(i)}\right) \quad \text{Eq. 13}$$

Figure 18 shows the PDF for different selections of database based on the episodes. Each episode has different control procedures characterized by the end of speed of PSP #2 $(w_2)_{end}$ and the trip temperature $t_{trip}$, and there are in total 1024 episodes. 1:10:100 means that the database divides a total 100 episode uniformly into 10 parts, while the first episode is selected from each part and assembled as a database. It can be found that the distribution of first 100 episodes is very different from the distribution of 1024 episodes. Meanwhile, the number of divisions does not significantly influence the PDF distributions.

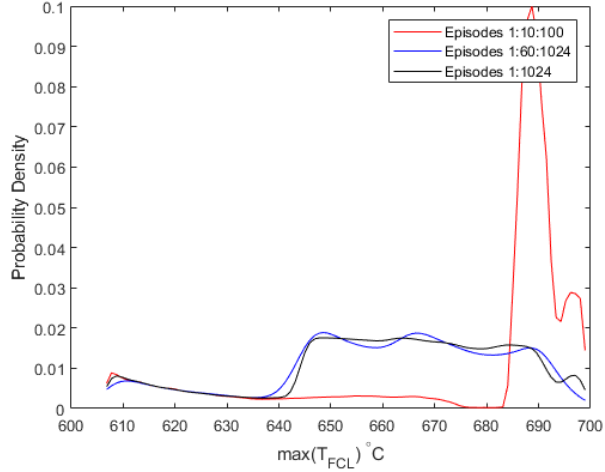

Figure 18: PDF distributions for different selections of database

Figure 19 plots the root mean squared error (RMSE), defined in Eq. 9, versus the quantified mutual information by symmetric K-L divergence. It is found that the DT-D testing errors are strongly correlated with the mutual-information between training and testing databases. The Pearson correlation coefficient (PCC) is calculated to determine the degree of linear correlation as in Eq. 14

$$\rho_{KL,RMSE} = \frac{cov(KL, RMSE)}{\sigma_{KL}\sigma_{RMSE}} \quad \text{Eq. 14}$$

where $cov(KL, RMSE)$ represents the covariance of K-L divergence and the testing RMSE; $\sigma_{KL}$ and $\sigma_{RMSE}$ represent the standard deviation of K-L divergence and testing RMSE respectively. The PCC value is found to be ~0.4, which again indicates a strong correlation.

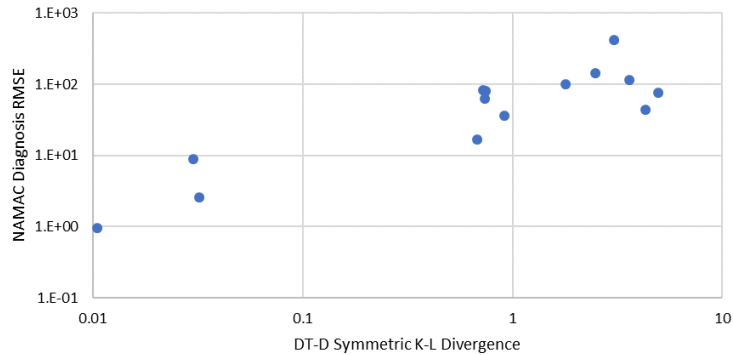

Figure 19: Diagnosis testing error with different coverages quantified by mutual information

### A.2. Training-Parameter Assessment for Digital Twin for Diagnosis

In this study, the selection of the target loss is assessed for the digital twin for diagnosis (DT-D). The target loss is used as one of the stop criteria that stop neural network training when the training mean squared error is lower than the selected value. Figure 20 shows the plot of training and testing errors with different target MSE. The issue space characterization for the testing database is the same as those for the training database – i.e. testing database is covered and globally interpolated by the training database. It is found that the target loss is reached as most training errors equal the target loss, which indicates the target loss is triggered to stop the training. It is also found that when the target loss is smaller than 0.5, the testing errors are generally larger than the training error, which indicates an overfitting with limited generalization capability in interpolated cases. When the target loss is larger than or equal to 0.5, the testing and training errors are almost the same.

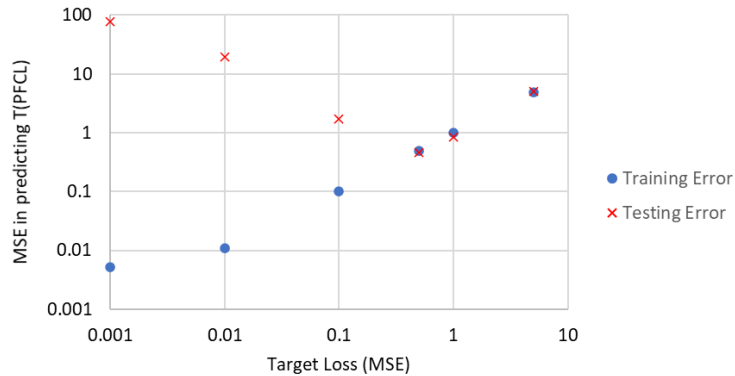

Figure 20: Plot of training and testing errors given different target MSE for DT-D. The testing database is covered by the training database (globally interpolated condition)

In addition to the covered and globally interpolated case, this study also evaluates the impact of target loss on DT-D's performance in globally extrapolated conditions. Databases from two extrapolated conditions are used: one with more severe primary pump malfunctions than the training case, while another one with lighter primary pump malfunctions than the training case. Meanwhile, an interpolated testing is performed with the same degree of pump malfunction as the training database. All three cases have the same control actions to mitigate the malfunctions. The red dashed line represents the acceptance criterion, where the RMSE is expected to be less than 30℃. It is found that the error of DT-D in the extrapolated conditions is generally larger than the error in the interpolated conditions. When the target loss is less than 0.5, the errors of testing in extrapolated conditions are larger than the acceptance criteria, which indicates a limited generalization capability and overfitting issue. When the target loss is larger than or equal to 0.5, the error of testing in extrapolated conditions become acceptable.

Both tests with globally interpolated and extrapolated conditions indicate that the DT-D errors are strongly correlated with the target loss of machine-learning training. Sensitivity analysis and optimization techniques are needed to properly select training parameters.

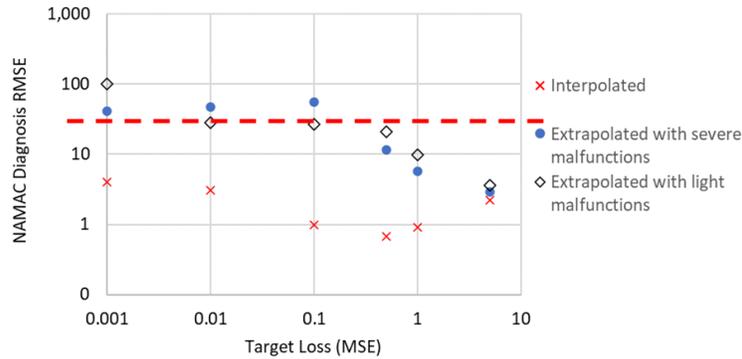

Figure 21: Plot of training and testing errors given different target MSE for DT-D. The testing database is not fully covered by the training database (globally extrapolated condition). The red dashed line represents the acceptance criterion for DT-D.

### A.3. Input-Variable Assessment for Digital Twin for Diagnosis

To evaluate the impact of input variables on the error of DT-D, this study designs two case studies: one missing input due to single sensor failure, distribution of error given all inputs along each transient. The case with missing inputs is designed by replacing one out of three DT-D input by not-a-number (NaN) index. In most of the neural network toolbox, these missing values can either be replaced by the average of other valid number of the same input type, or they can be removed. Since DT-D is required to continuously infer the peak fuel centerline temperature, this study replaces all NaN value by the average of other valid number. Figure 22 shows the inputs for DT-D with normal and failed sensor that measures the upper plenum sodium temperature.

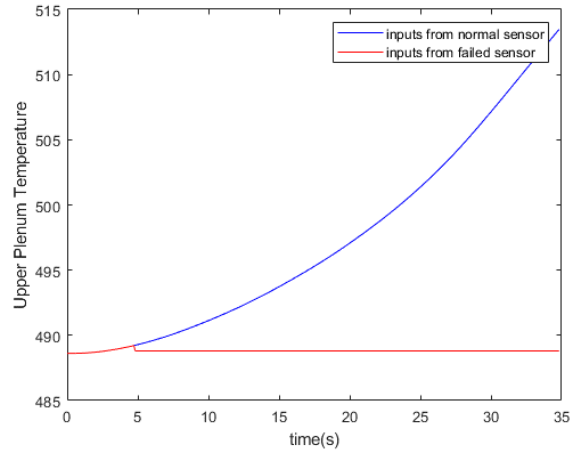

Figure 22: Comparison of inputs DT-D with normal sensor and failed sensor for upper plenum temperature. The sensor failure starts from 5 sec.

Figure 23 compares the DT-D output for the fuel centerline temperature $T_{PFCL}$ given normal sensors and a single sensor failure. The real value from the database overlaps with the predictions by normal sensors. It is found that the DT-D responses differently to sensor failures at different locations. Diagnosis outputs are more sensitive to sensor readings at high-pressure plenum than others, and the prediction start to fluctuate drastically starting from 15sec, which is 10sec after sensor failures. For failures of other sensors, the prediction starts to fluctuate from 25sec, which is 20 sec after the sensor failure, with abnormal sensors at low-pressure or abnormal upper plenum

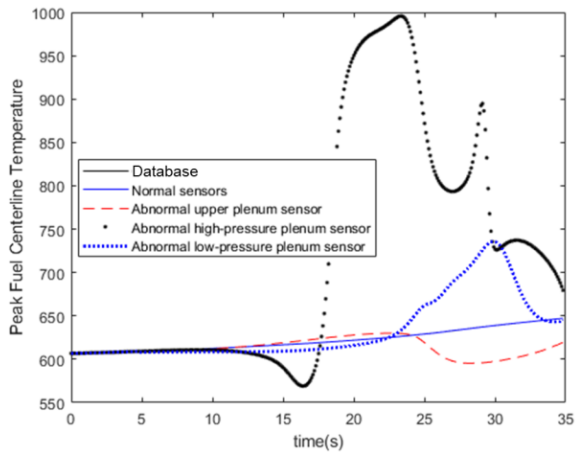

Figure 23: Comparison of predicted $T_{PFCL}$ given normal sensors and a single sensor failure

As an average value, MSE cannot clearly indicate the distribution of errors along one transient. To better demonstrate how DT-D makes predictions along different episodes, this study evaluates the distribution of errors for all 1024 episodes, while each episode contains a 2000-step transient. Each step is 0.1sec, and the entire transient lasts for 200sec. Figure 24 shows the surface plot for DT-D predictions error, quantified by root mean squared error (RMSE), for all 1024 episodes. The maximum RMSE is found to be 6.7℃, which is acceptable comparing to the 30℃ criterion.

However, it is found that the prediction errors suddenly increase at 35sec and gradually decrease after then.

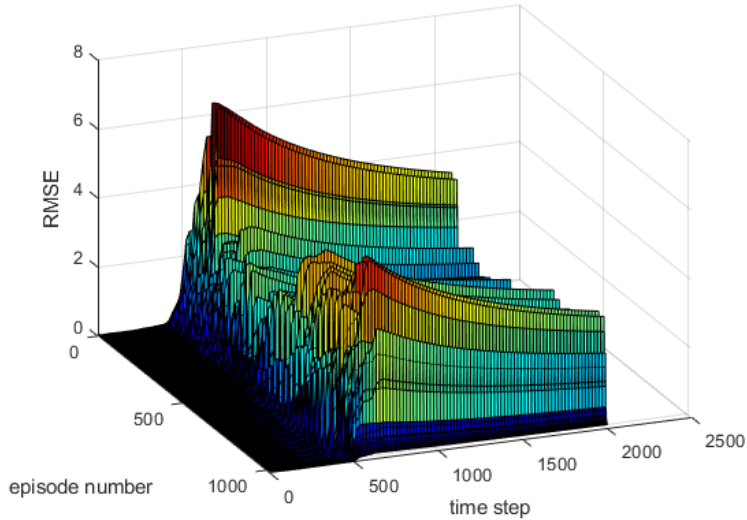

Figure 24: Surface plot for DT-D prediction error, quantified by RMSE, for all 1024 episodes, while each episode contains a 200-step transient

### A.4. Coverage and Training-Parameter Assessment for Digital Twin for Prognosis

Similar to the DT-D assessment, the impact of coverage is evaluated for the digital twin for prognosis (DT-P). Figure 25 compares the predicted outputs against real values for testing in extrapolated conditions. It is found that the DT-P cannot correctly predict the consequence factor in extrapolated conditions. Compared to the training results in Figure 10, the DT-P error is strongly affected by the coverage condition.

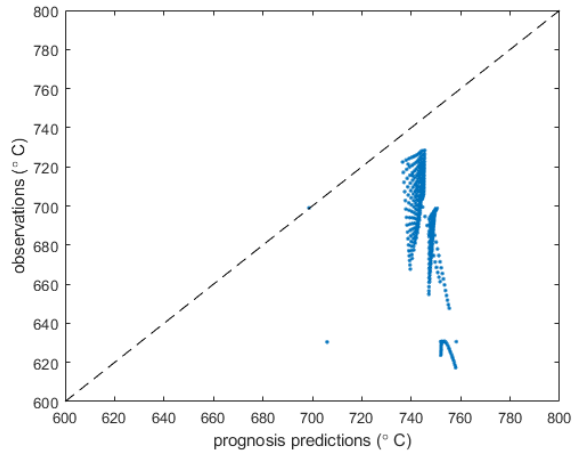

Figure 25: Comparison of predicted outputs against real values from databases for DT-P testing. Dash lines are diagonal lines

At the same time, the impact of a training parameter – e.g. target training loss – on the DT-P error is evaluated. It is found that the target loss is reached after the training as most training errors equal the target loss, which indicates the target loss is triggered to stop the training. However, the DT-P error exceeds the acceptance criterion (30°C) for all selections of target loss. As a result, the training errors for DT-P is correlated with the target loss. However, in extrapolated conditions, the impact from the target loss is limited.

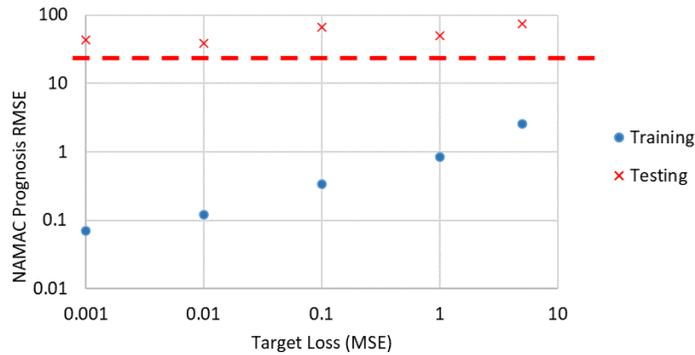

Figure 26: Plot of training and testing errors given different target MSE for DT-P. The testing database is not fully covered by the training database (globally extrapolated condition). The red dashed line represents the acceptance criterion for DT-P.